\documentclass[11pt,a4paper]{article}
\usepackage{amsmath,amssymb}
\usepackage{yfonts}
\usepackage{bbm}

\def\hybrid{\topmargin -20pt    \oddsidemargin 0pt
        \headheight 0pt \headsep 0pt
        \textwidth 6.25in       
        \textheight 9 in       
        \marginparwidth .875in
        \parskip 5pt plus 1pt
          \jot = 1.5ex
   }
\hybrid
\numberwithin{equation}{section}
\numberwithin{table}{section}

\newcommand{\hh}{\textgoth{h}}
\newcommand{\be}{\begin{equation}}
\newcommand{\ee}{\end{equation}}
\newcommand{\bea}{\begin{eqnarray}}
\newcommand{\eea}{\end{eqnarray}}

\begin{document}
\thispagestyle{empty}
\baselineskip=14pt
\parskip 5pt plus 1pt

\vspace*{-1.5cm}
\begin{flushright}    
  {\small
  }
\end{flushright}

\vspace{2cm}
\begin{center}        
  {\LARGE \bf M-theory Compactifications to Three Dimensions\vspace{0.3cm}\\ with M2-brane Potentials}
\end{center}

\vspace{0.75cm}
\begin{center}        
  {\bf Cezar Condeescu$^{1,2}$, Andrei Micu$^{2}$, Eran Palti$^{3}$}
\end{center}

\vspace{0.15cm}
\begin{center}        
  \emph{$^{1}$ INFN, Sezione di Roma ``Tor Vergata",\\
             Via della Ricerca Scientifica 1, 00133 Roma, Italy}
             \\[0.15cm]
  \emph{$^{2}$ Department of Theoretical Physics, IFIN-HH, \\
               Reactorului 30, 077125, Magurele/Ilfov, Romania}
             \\[0.15cm]
  \emph{$^{3}$ Institut f\"ur Theoretische Physik, Ruprecht-Karls-Universit\"at, \\
             Philosophenweg 19, 69120, Heidelberg, Germany}
             \\[0.15cm]

\end{center}

\vspace{2cm}


\begin{abstract}
  \vspace{0.5cm} We study a class of compactifications of M-theory to
  three dimensions that preserve $N=2$ supersymmetry and which have
  the defining feature that a probe space-time filling M2 brane feels
  a non-trivial potential on the internal manifold. Using M-theory/F-theory duality such
  compactifications include the uplifts of 4-dimensional $N=1$ type
  IIB compactifications with D3 potentials to strong coupling. We
  study the most general 8-dimensional manifolds supporting these
  properties, derive the most general flux that induces an M2
  potential, and show that it is parameterised in terms of two real
  vectors. We study the supersymmetry equations when only this flux is
  present and show that over the locus where the M2 potential is
  non-vanishing the background takes the form of a Calabi-Yau
  three-fold fibered over a 2-dimensional base spanned by the flux
  vectors, while at the minima of the potential the flux
  vanishes. Allowing also for non-vanishing four-form flux with one
  leg in the internal directions we find that the Calabi-Yau
  three-fold in the fibration is replaced by an $SU(3)$-structure
  manifold with torsion classes satisfying $2 W_4=-W_5$.

\end{abstract}
\vspace{1.7cm}

\begin{itemize}
{\small
\item[E-mails:] {\tt condeescu@roma2.infn.it}\,;\ {\tt amicu@theory.nipne.ro}\,;
 {\tt palti@thphys.uni-heidelberg.de}.}
\end{itemize}

\clearpage


\newpage

\setcounter{page}{1}
\tableofcontents

\section{Introduction}

In this paper we study compactifications of M-theory to three dimensions that preserve $N=2$ supersymmetry and which induce a potential for space-filling M2-branes. Though interesting as three-dimensional vacua in themselves, our primary motivation for studying them originates in F-theory \cite{Vafa:1996xn}. Four-dimensional $N=1$ F-theory vacua can be defined as dual to a particular limit of $N=2$ M-theory compactifications to three-dimensions. This definition relies on the assumption that the 8-dimensional manifold on which M-theory is compactified is elliptically fibered. Then the appropriate limit where four-dimensional physics is recovered is that of a vanishing fibre. This duality is most often used to construct F-theory backgrounds as dual to M-theory compactifications on elliptically fibered Calabi-Yau manifolds.\footnote{Though see recent work on manifolds with $Spin(7)$ holonomy \cite{Bonetti:2013fma,Bonetti:2013nka}.} Such compactifications can exhibit realistic particle physics models and have been under intensive study in recent years \cite{Heckman:2010bq,Weigand:2010wm,Maharana:2012tu}. A well-understood generalisation of such models is to include a particular $(2,2)$ background four-form flux on the M-theory side, which is dual to both background closed-string and brane fluxes on the F-theory side. Such fluxes play a crucial role in moduli stabilisation and in generating chirality in particle physics models. It is known that the backreaction of such fluxes deforms the background only to conformal CY \cite{Becker:1996gj,Giddings:2001yu}, so that the relevant compactifications are warped $N=2$ compactifications of M-theory to three dimensions. This class of compactifications have the property that space-filling probe M2-branes on the M-theory side, which are dual to space-filling probe D3-branes on the F-theory side, are BPS at all points of the (conformal) CY and therefore feel no potential \cite{Giddings:2001yu}.

There are a number of interesting departures from such backgrounds which can be characterised by the fact that a potential is generated for (probe) D3-branes. The most familiar are backgrounds of type IIB string theory that support gaugino condensation on D7-branes, or certain types of D3-instantons that contribute to the superpotential, as used for moduli stabilisation in the KKLT and Large volume scenarios \cite{Kachru:2003aw,Balasubramanian:2005zx}. The fact that these induce a potential for D3-branes can be shown by performing a 1-loop string calculation \cite{Berg:2004ek} or through gravitational back-reaction \cite{Baumann:2006th}. As well as being used for moduli stabilisation, there is a direct cosmological use for type IIB backgrounds that induce D3-potentials for inflation \cite{Dasgupta:2002ew,Baumann:2009ni}. Such backgrounds also play an important role in particle physics model building. In F-theory models where all the generations are realised on a single matter curve the Yukawa coupling at a point of intersection is rank one \cite{Cecotti:2009zf,Conlon:2009qq} unless an appropriate deformation of the geometry which induces a D3 potential by flux \cite{Cecotti:2009zf} or non-perturbative effects \cite{Marchesano:2009rz} is present.\footnote{Interestingly, such backgrounds also induce non-commutative deformations on the world-volume theories of the 7-branes \cite{Dasgupta:2002ew,Cecotti:2009zf,Marchesano:2009rz}.} Since all the listed interesting IIB/F-theory backgrounds have potentials for D3-branes they are not within the better understood class of warped CY compactifications. Although an approach of neglecting the backreaction of the effects that induced the D3 potential on the geometry and continuing to use a CY background may be a valid approximation for some purposes, an understanding of the backreaction is essential for applications where the effects are large or when a treatment of the background in a 10-dimensional sense, rather than a 4-dimensional effective theory, is important. Studying such backgrounds, particularly at strong coupling, using F-theory/M-theory duality should therefore involve some understanding of three-dimensional M-theory backgrounds which preserve $N=2$ supersymmetry and have a potential for M2-branes.

Having identified the presence of a D3/M2 brane potential as the defining feature we are interested in, it is essential to understand what are the properties of the background geometry and flux which induce such a potential. There is a rather general and neat answer to this question in type IIB supergravity. The key property of the background is the structure group of the metric once any flux/branes are back-reacted.
Recall that a nowhere vanishing spinor on a manifold reduces its structure group. Since the presence of a nowhere vanishing internal spinor is a direct requirement for the background to preserve some supersymmetry a reduced structure group typically characterises supersymmetric backgrounds  \cite{Gauntlett:2002sc}. In particular the most general type IIB (supergravity) backgrounds which preserve four-dimensional $N=1$ supersymmetry have an $SU(3)\times SU(3)$ structure group \cite{Grana:2005sn,Grana:2005jc}. Here each $SU(3)$ is associated to a spinor on the manifold, but the two spinors are not everywhere orthogonal (which would lead to $SU(2)$-structure) or parallel (leading to $SU(3)$-structure) but rather the angle between them varies over the manifold. Now an interesting result of \cite{Martucci:2006ij,Koerber:2006hh} is that backgrounds which support potentials for D3-branes are those which have such an $SU(3)\times SU(3)$ structure group, and further, that the minimum for the potential, where the D3-branes are BPS, occurs exactly on the loci where the two spinors become parallel, yielding a `local' $SU(3)$-structure.\footnote{We will use the notion of a `local' structure group often in the paper. We define it as the structure group that would result should the properties of the spinors on a local submanifold where the local structure group is defined be extended to the full space.} This general result was checked for the particular case of D7 gaugino condensation in \cite{Koerber:2007xk} which confirmed that their backreaction indeed changes the structure group from $SU(3)$ to $SU(3)\times SU(3)$. It was also confirmed to some extent in \cite{Baumann:2009qx,Baumann:2010sx} for the case of D3-potentials induced by non-Imaginary-Self-Dual background flux, specifically by showing that the backreaction of D7 gaugino-condensation can locally be viewed as such a background flux.

One aim of this paper is to develop analogous relations between structure groups and M2-brane potentials in M-theory. The relation between structure groups and non-vanishing spinors in 8 dimensions is rather different from the more familiar 6 and 7 dimensions. Consider a compactification on an 8-dimensional manifold $X_8$. A requirement for preserving some supersymmetry in 3-dimensions is the existence of a nowhere vanishing Majorana spinor on $X_8$. This is so that the 11-dimensional M-theory supersymmetry spinor decomposes into a product of a 3-dimensional and 8-dimensional Majorana spinor. However, the existence of a nowhere vanishing Majorana spinor in 8 dimensions does not imply a reduction of the structure group.\footnote{The stabiliser group of a Majorana spinor is $G_2$. However we use the notion of a global structure group as maximal over the manifold, while the $G_2$ stabiliser may enhance over certain loci.} This only occurs in the presence of nowhere vanishing Majorana-Weyl spinors: a single such spinor implies $Spin(7)$-structure, while two such spinors imply $G_2$ or $SU(4)$-structure if they have the opposite or same relative chirality respectively. Now the number of supersymmetries preserved by an M2-brane in a background is given by the number of independent covariantly constant Majorana-Weyl spinors of fixed chirality \cite{Becker:1995kb}. Since the manifolds with fixed structure group have a fixed number of Majorana-Weyl spinors of fixed chiralities M2-branes are BPS over the whole space and preserve the supersymmetries of the background. Therefore a connection between structure groups and M2 potentials for 8-dimensional manifolds is not obvious.

There is a useful way to think about the M2-potential in terms of 8-dimensional local structure groups, i.e. submanifolds of $X_8$ over which the spinors satisfy certain properties such as being non-vanishing. The two nowhere vanishing Majorana spinors can be decomposed into 4 Majorana-Weyl spinors, but any of
these four may vanish on certain loci. In the generic point on $X_8$ all four are non-vanishing and we have a local $SU(3)$-structure on
$X_8$, over certain loci one of the Majorana-Weyl components in each Majorana spinor may vanish so that we are left with two Majorana-Weyl
spinors of same or opposite chirality leading to local $SU(4)$ or $G_2$-structures. M2-branes are calibrated by a Majorana-Weyl spinor
of fixed chirality and therefore on the generic $SU(3)$-structure loci they preserve no supersymmetry, on $G_2$ loci they preserve $N=1$
supersymmetry and on $SU(4)$ loci they preserve $N=2$ supersymmetry. They therefore feel a potential in such backgrounds
with minima at $SU(4)$ and $G_2$-structure loci. Such backgrounds are therefore of interest following our original motivation and much of
this paper is dedicated to exploring their properties.

In \cite{Tsimpis:2005kj} it was shown that one can induce a connection between the existence of a Majorana spinor on $X_8$ and global G-structures by defining an auxiliary 9-dimensional manifold, $Y_9$, which is just the product of the 8-dimensional one with a circle $Y_9 = X_8 \times S_1$. Now the existence of an 8-dimensional nowhere vanishing Majorana spinor on $X_8$ induces a nowhere vanishing Majorana spinor on $Y_9$ and this is known to imply a reduction of the structure group of $Y_9$ to $Spin(7)$. Therefore it is quite natural to work with this auxiliary 9-dimensional manifold when studying the supersymmetry properties of the background. In this paper we are interested in $N=2$ vacua and so require two covariantly constant, and therefore nowhere vanishing, Majorana spinors. The requirement of Majorana spinors rather than Majorana-Weyl spinors leads to the most general 8-dimensional backgrounds that preserves $N=2$ in three dimensions from M-theory, in this sense they are the analogs of $SU(3)\times SU(3)$ structure 6-dimensional backgrounds in IIB. As with the case of one spinor, the nowhere vanishing Majorana spinors do not induce a reduction of the structure group in 8 dimensions. It is possible to consider again an uplift to $Y_9$, but although the structure group is reduced to at least $Spin(7)$, there is in general no further reduction on $Y_9$ due to the second spinor. Nonetheless, the 9-dimensional approach is useful for treating the 8-dimensional local structure groups in a unified way and we will utilise it in this work.

So far we have discussed only the geometry part of the compactification and not the energy-momentum tensor that sources it. In this work we will study the background flux that can source M2-brane potentials. In relation to the previous discussion of physics sources for a D3-potential in IIB, this flux can be thought of either as non-trivial background flux or, in the spirit of \cite{Baumann:2009qx,Baumann:2010sx}, as flux that is accounting for the backreaction of localised sources. We will be able to give the form of the most general flux that generates an M2-potential in terms of 8-dimensional $SU(3)$-structure geometric objects and two real one-forms that parameterise the flux. For the simple case where only this type of flux and four-form flux with one leg in the internal directions, which we henceforth refer to as 1-form flux, are present the supersymmetry equations simplify considerably and we are able to present them as differential relations on the $SU(3)$-structure forms and extract some key properties. We find that for this limited flux configuration the compactification must be to 3-dimensional Minkowski space. If we further impose the vanishing of the 1-form flux we find that the M2-potential is only along the directions parameterised by the two singlet vectors of the 8-dimensional $SU(3)$-structure. The torsion classes are such that on the generic $SU(3)$-structure locus the manifold can be described as a 6-dimensional Calabi-Yau fibered over a 2-dimensional base which is spanned by the singlet vectors, and over which the M2-branes have a non-trivial potential. While over the special $SU(4)$ and $G_2$-structure loci the flux vanishes. The more involved background where we also allow for non-vanishing 1-form flux leads to a similar configuration but the 6-dimensional fibre is not Calabi-Yau but has non-vanishing torsion classes (which satisfy the relation $2W_4=-W_5$). We will also present the supersymmetry equations in differential form for the most general flux configurations that have other, non M2-potential inducing, fluxes turned on. Though this substantially more complicated system is difficult to analyse in as much detail.

The outline of the paper is as follows. In section \ref{sec:n2d3g} we study the geometric properties of the background using G-structures. In particular we describe 8-dimensional manifolds with varying structure groups and M2-potentials. In section \ref{sec:susycond} we study the supersymmetry equations and formulate them in a way compatible with the 9-dimensional geometry. In particular we identify the flux which is responsible for the M2-potentials, and parameterise it in terms of $SU(3)$-structure objects. In section \ref{sec:anapartfl} we study the implication of the supersymmetry equations for backgrounds supporting such a flux. We summarise our results in section \ref{sec:summary}.

\paragraph{Note 1:} The results presented in this paper rely on quite
lengthy calculations. Although most of these calculations can be done
by hand we made extensive use of symbolic calculation applications
which are able to manipulate tensors and/or gamma matrices, in order to check and derive
some of our results. We acknowledge the use of the following
resources: Mathematica \cite{Math}, MathTensor \cite{MathT}, Cadabra,
\cite{Peeters:2007wn,Peeters:2006kp}, Gamma \cite{Gran:2001yh} and
xTensor \cite{XT}.

\paragraph{Note 2:} This paper has some overlap with a project which was
initiated together with M.~Babalic, I.~Coman and C.~Lazaroiu to whom
we thank for insights in the subject of M-theory compactifications to three dimensions.

\section{N=2, D=3 M-theory compactifications and G-structures}
\label{sec:n2d3g}

\subsection{Supersymmetric compactification of M-theory}
\label{sec:sucompm}

The effective action of M-theory is described by eleven dimensional supergravity consisting of the following fields: metric $g_{MN}$, three-form potential $C$ with corresponding field strength $G=dC$ and gravitino $\Psi_M$. The action can then be written in the following way \cite{Cremmer:1978km}
\begin{equation}
S_{11}=\frac{1}{2}\int d^{11}x \sqrt{-g} \left(R-\frac{1}{2}G\wedge *G - \frac{1}{6}C \wedge G \wedge G\right) \;.
\end{equation}
We shall be interested in supersymmetric flux backgrounds. They correspond to fluxes for which the background gravitino vanishes together with its supersymmetry variation with 11-dimensional (Majorana) spinor parameter $\epsilon$
\begin{equation}
\delta \Psi_M = \nabla_M \epsilon - \frac{1}{288}\left(\Gamma_M{}^{NPQR} - 8 \delta^N_M \Gamma^{PQR}\right) G_{NPQR} \epsilon = 0 \;.
\label{susyvar}
\end{equation}
The matrices $\Gamma_M$ are taken to satisfy the eleven-dimensional Clifford algebra with metric signature $(-,+,...,+)$.

The supersymmetry equations should be supplemented by the Bianchi identity and equations of motion for $G$ which read\footnote{It is usually the case that the equations of motion and the supersymmetry equations imply the Einstein equations, though we will not prove it here for the particular class of backgrounds under consideration.}
\begin{align}
dG &= 0 \;, \\
d\star G &= - \frac12 G \wedge G  + 2 \pi T_6 X_8 \;. \label{eomG}
\end{align}
The last term in (\ref{eomG}) corresponds to a higher derivative gravitation correction \cite{Duff:1995wd}, where $T_6$ is the M5-brane tension and $X_8$ is a known combination of the first and second Pontrjagin forms. This correction is important as it allows the support of solutions to three-dimensional Minkowski space in the presence of background flux for compact smooth manifolds. Equation (\ref{eomG}) is usually imposed as integrated over the manifold in which case the first term vanishes and the last term gives the Euler number of the manifold, This leads to the familiar D3/M2 tadpole cancellation constraint.\footnote{Note that for manifolds with 8-dimensional $G_2$-structure the Euler number is forced to vanish \cite{Isham:1987qe}.}

The compactification Ansatz is chosen by imposing a 3-8 split of the
11 dimensional manifold. We choose a metric which is a warped product
of 3-dimensional space-time and an 8-dimensional Euclidean manifold.
\begin{equation}
  ds_{11}^2 = e^{2 \Delta} (ds_{2,1}^2 + ds_8^2) \;.
  \label{metric}
\end{equation}
The 11-dimensional index $M$ decomposes into an external 3-dimensional index $\mu=1,2,3$, and an internal 8-dimensional index $\alpha=1,..,8$.
For the 4-form field strength $G$ we choose the most general ansatz compatible with Lorentz invariance
\begin{equation}
  G = e^{3\Delta}\left( \tilde{f} \wedge \mathrm{Vol}_3 + F \right) \;,
  \label{flux}
\end{equation}
where $\tilde{f}$ is a one-form and $F$ is a 4-form on the internal manifold, while $\mathrm{Vol}_3$ is the volume element of the external space-time.

The eleven-dimensional Clifford algebra is decomposed according to the following equations
\begin{align}
\Gamma_\mu & = e^\Delta (\gamma_\mu \otimes \gamma_9) \nonumber \;,\\
\Gamma_\alpha & = e^\Delta (\mathbbm{1} \otimes \gamma_\alpha) \;, \label{cliff}
\end{align}
with the $2 \times 2$ matrices $\{\gamma_\mu\ ;\  \mu=1,2,3\}$ generating the three-dimensional Clifford algebra $Cl(2,1)$. An explicit representation can be given in terms of the Pauli matrices. The $16\times 16$ matrices $\gamma_\alpha$ are taken to be real and symmetric. They generate the eight-dimensional Clifford algebra $Cl(8,0)$.

We can decompose the 11-dimensional supersymmetry parameter $\epsilon$, which is an 11-dimensional Majorana spinor, according to the 3-8 split as
\begin{equation}
  \label{spdec}
  \epsilon = e^{-\tfrac{\Delta}2}\eta \otimes \xi \; ,
\end{equation}
where $\eta$ is a 3-dimensional Majorana spinor, while $\xi$ is an 8-dimensional Majorana
spinor. Each non-vanishing spinor $\xi$ defines by the relation above
one spinor $\eta$ in three dimensions. Therefore for $N=2$ supersymmetry we
need two spinors $\xi_i$ on the internal manifold. Note that in 8 dimensions there
exist Majorana-Weyl spinors. However we do not impose any chirality
condition on the internal spinors as from the supersymmetry equation
it is clear that only the Majorana condition is necessary. Imposing the
Weyl property is an additional constraint. In most studies of M-theory
compactifications to 3 dimensions so far the Majorana-Weyl condition was
imposed for simplicity but, as emphesised in
\cite{Martelli:2003ki,Tsimpis:2005kj}, this is not the most general case.

\subsection{General 8-dimensional manifolds preserving $N=2$ supersymmetry}
\label{sec:susy8d}

In this section we present a detailed description of the manifolds on which
we compactify M-theory. As explained before, such manifolds admit two
independent, nowhere vanishing Majorana spinors $\xi_{1,2}$. We shall see further
that the supersymmetry conditions imply that the norm of these spinors
is constant \cite{Martelli:2003ki} and therefore without loss of
generality we shall assume that the two spinors are orthonormal
\begin{equation}
  \label{orthonorm}
  \xi_i^T \xi_j = \delta_{ij}\;\;,\;\;\; i,j = 1,2\; .
\end{equation}
Since in 8 dimensions the Majorana and Weyl conditions are compatible,
we can split the two spinors into spinors of definite chirality
\begin{equation}
  \label{MWdec}
  \xi_i = (\xi_+)_i + (\xi_-)_i \; .
\end{equation}
However, the Majorana-Weyl components $(\xi_\pm)_i$ are no longer
required to have constant norm and moreover they can even vanish at
certain points. The only requirement is the unit norm of the spinors
$\xi_i$
\begin{equation}
  ||(\xi_+)_i||^2 + ||(\xi_-)_i||^2 = 1\; .
\end{equation}
In the case that all Majorana-Weyl components are everywhere non-vanishing we are
actually dealing with an 8-dimensional manifold with $SU(3)$ structure which
preserves (a maximum of) $N=4$ supersymmetry in 3 dimensions. If some of the
Majorana-Weyl components vanish identically over the entire internal
manifold, while the others are non-vanishing, then we are dealing with
manifolds with $G_2$ structure or manifolds with $SU(4)$ structure
depending on the relative chirality of the non-vanishing spinors.
  We see that the fact that the Majorana-Weyl components are allowed to
vanish at certain points implies that such manifolds do not admit a
global reduction of the structure group. At generic points they look
like $SU(3)$ structure manifolds, while at special points they look like
$G_2$ or $SU(4)$ structure manifolds.\footnote{It would be interesting to explore connections between these backgrounds and $SU(4)\times SU(4)$ backgrounds as studied in \cite{Prins:2013koa,Rosa:2013lwa}.}

Note that we did not consider the case where only one of the four
  Majorana-Weyl components vanishes. This is because on an 8-dimensional manifold once we are given
three linearly independent, non-vanishing, Majorana--Weyl spinors
which are not of the same chirality, one can define a fourth spinor
such that we end up with two spinors of one chirality and two of the
other.
To prove this, suppose that we have an
eight-dimensional manifold which has
two spinors, $\xi_1$ and
$\xi_2$ of positive chirality and one, $\chi_1$ of negative
chirality. The two spinors of positive chirality define a $SU(4)$
structure and therefore one finds an almost complex structure which is
given by
\begin{equation}
  J_{\alpha \beta} = \xi_1^T \gamma_{\alpha \beta} \xi_2
\end{equation}
We can define a fourth spinor $\chi_2$ which is of negative chirality
\begin{equation}
  \chi_2 = \tfrac12 J^{\alpha \beta} \gamma_{\alpha \beta} \chi_1 \; ,
\end{equation}
and which has non-vanishing norm. This spinor is obviously orthogonal
to $\xi_{1,2}$ and because the matrix $\gamma_{\alpha \beta}$ is
antisymmetric in its spinorial indices, it is also orthogonal to
$\chi_1$. Therefore, in this way we have 4 non-vanishing Majorana-Weyl
spinors, two of positive and two of negative chirality, which define a
$SU(3)$ structure in 8 dimensions.

\subsubsection{9-dimensional uplifts to $Y_9 = X_8 \times S_1$}
\label{sec:9dg2}

It was pointed out in \cite{Tsimpis:2005kj} that we can associate the existence of a nowhere vanishing Majorana spinor on $X_8$ to a reduced structure group by uplifting it to an auxiliary 9-dimensional manifold, $Y_9$, defined as the direct product of the 8-dimensional manifold under consideration $X_8$ and a circle $Y_9 = X_8 \times S_1$. For the case of 2 Majorana spinors this procedure is not as effective since, as we will show, the structure group does not reduce further on $Y_9$. However, uplifting to 9-dimensions is still a useful procedure because it will allow us to describe the local structure groups of $X_8$ in a unified way. In particular the 8-dimensional structure groups, and the related M2 potential, will be mapped to an angle between two 9-dimensional vectors.

We therefore go on to study 9-dimensional manifolds supporting two nowhere vanishing Majorana spinors. It is most common to study such manifolds from the perspective of spinor bilinears which can be constructed out of the spinors. Since 9-dimensional Euclidean gamma
matrices can be chosen real (and therefore symmetric) the only spinor bilinears which can be defined are the following
\begin{align}
  \label{bilinears}
  (V_1)_m & =  \xi_1^T \gamma_m \xi_1 \; , \qquad   (V_2)_m = \xi_2^T
  \gamma_m \xi_2 \; , \qquad   (V_3)_m = \xi_1^T \gamma_m \xi_2 =
  \xi_2^T \gamma_m \xi_1 \; , \nonumber\\
  K_{mn} & =  \xi_1^T \gamma_{mn} \xi_2 = - \xi_2^T \gamma_{mn} \xi_1
  \; , \qquad
  \Psi_{mnp} = \xi_1^T \gamma_{mnp} \xi_2 = - \xi_2^T \gamma_{mnp}
  \xi_1 \; , \nonumber\\
  (\Phi_1)_{mnpq} & =  \xi_1^T \gamma_{mnpq} \xi_1 \; , \qquad
  (\Phi_2)_{mnpq} = \xi_2^T \gamma_{mnpq} \xi_2 \; , \qquad
  (\Phi_3)_{mnpq} = \xi_1^T \gamma_{mnpq} \xi_2 \;.
\end{align}
Here $\gamma_m$ are 9-dimensional gamma matrices, or generators of $Cl(9,0)$, with $m=1,...,9$.
These forms are not independent as certain products of such bilinears
can be expressed in terms of linear combinations of these
bilinears. The complete set of relations which the forms above satisfy
is given in appendix \ref{sec:frz}.

For an efficient description of such manifolds it is important to know
the number of independent vectors. The forms defined
above include three vectors. The Fierz
relations for the vectors imply
\begin{align}
  ||V_1||^2 & =  ||V_2||^2 = 1\; , \qquad ||V_3||^2 = \frac12 \left(1-\alpha\right) \;,\; \\
  V_1 \cdot V_2 & \equiv  \alpha \; , \qquad V_1\cdot V_3 = V_2\cdot V_3 = 0 \; .
\end{align}
Here we define the usual contraction $V \cdot U=V^m U_m$.  Therefore
the number of independent vectors is governed by a real parameter
$\alpha$ which is the scalar product of the vectors $V_1$ and
$V_2$. Since the vectors $V_1$ and $V_2$ are of unit norm, this
parameter $\alpha$ is in fact the cosine of the angle between these
vectors and therefore can take values in the interval $[-1, 1]$. At
generic values of this parameter, all the three vectors are
independent and of non-vanishing norm. This is the case of a local $SU(3)$ structure. We denote this a `local structure' because $\alpha$ varies over $X_8$ and so
particular values of it define certain submanifolds. Of course the notion of structure group only has a global meaning.
However the terminology of a local structure is well defined by the properties of the spinors over the submanifold and will be used extensively in this work.
For $\alpha=-1$ the vectors $V_1$ and $V_2$ are no longer
independent (they are anti-parallel) while $V_3$ has unit
norm. Therefore we are left with two independent unit vectors. This is
the case of a $G_2$ structure. Finally, if $\alpha = 1$, the vectors
$V_1$ and $V_2$ are parallel while $V_3$ vanishes. This is the case of
an $SU(4)$ structure. These local 9-dimensional structure groups dictated by the parameter $\alpha$ are directly inherited by the 8-dimensional manifold. In particular the physics we are interested in, which is the potential
for a probe space-filling M2-brane, is therefore directly related to the
variation of $\alpha$ over the internal manifold. We will show this in more detail in section \ref{sec:m2pot}.

Before going into more details about these manifolds it will be useful to understand the special cases of $\alpha= \pm 1$.

\subsubsection{Loci of $G_2$ structure: $\alpha=-1$}

On such loci, the Fierz relations presented in appendix simplify and we
find
\begin{align}
  \label{G2Fierz}
  V_1 & =  - V_2 = V \; , \qquad ||V||^2 = ||V_3||^2 = 1 \; , \\
  K & =  V \wedge V_3 \;,\\
  V \lrcorner \Psi & =  V_3 \lrcorner \Psi = 0  \;, \\
  \Phi_1 & =  -V_3 \wedge \Psi - *(V \wedge V_3 \wedge \Psi) \;,\\
  \Phi_2 & =  V_3 \wedge \Psi - *(V \wedge V_3 \wedge \Psi) \;,\\
  \Phi_3 & =  V \wedge \Psi \;.
\end{align}
This description looks much like a $G_2$ structure in 7-dimensions (which is
given in terms of the 3-form $\Psi$) and two additional vectors which
lift this $G_2$ structure to 9-dimensions. This is expected from the decomposition of the fundamental of $SO(9)$ under $G_2$
\be
{\bf 9} \rightarrow {\bf 7} \oplus {\bf 1} \oplus {\bf 1} \;. \label{so9g2}
\ee
A linear combination of these vectors gives the direction along the auxiliary circle used to uplift to 9-dimensions.

\subsubsection{Loci of $SU(4)$-structure: $\alpha=+1$}

The Fierz relations for the case $\alpha=1$ read
\begin{align}
  \label{SU4Fierz}
  V_1 & = V_2 = V \; , & ||V||^2 & = 1\; , & ||V_3||^2 & = 0 \; ,
  \\
 \Psi & = K \wedge V \; , & \Phi_+ & = -K \wedge K \; , &  K_{m[n} \left(\Phi_-\right)^m_{\;\;\;pq]} &= 2\left(\Phi_3\right)_{npq} \;,
\end{align}
where we define $\Phi_{\pm} = \Phi_1 \pm \Phi_2$.
Moreover it can be shown that when restricted to the subspace which is
orthogonal to $V$, $K$ is an almost complex structure and it is clear
that we can organise this orthogonal space as a space of SU(4)
structure with $\Phi_-$ and $2\Phi_3$ playing the role of the real and imaginary parts of the complex four-form.
The additional vector field should be understood
as the direction which we added to go to the auxiliary
nine-dimensional manifold.

\subsection{Parametrisation in terms of a $SU(3)$-structure}

The most interesting case is of when the angle $\alpha$, between the
vectors $V_1$ and $V_2$ varies.
This occurs over the generic patch on the manifold which manifests a local $SU(3)$-structure.
Over this locus we have that $\alpha \neq \pm 1$, and we will assume this in our present analysis and return to the limit points later.
In dealing with the more complicated $SU(3)$-structure case we are guided by the idea that, analogous to the $SU(4)$ and $G_2$-structure loci,
we expect to be able to describe it in terms of an (uplifted) 6-dimensional $SU(3)$-structure.
In order to unveil this structure let us first define
\begin{equation}
  \label{Vpm}
  V_{\pm} = V_1 \pm V_2 \; .
\end{equation}
Clearly $V_+$ and $V_-$ are mutually orthogonal and, since both $V_1$
and $V_2$ are orthogonal to $V_3$, they are also orthogonal to
$V_3$. However, these vectors no longer have unit norm but we find
\begin{equation}
  \label{normVpm}
  ||V_\pm||^2 = 2 (1 \pm \alpha) \; .
\end{equation}
The next step is to decompose all the forms in \eqref{bilinears} into
forms of lower or equal rank forms which are orthogonal to these
vectors. Let us use as an example the decomposition of $K$. We write
\begin{equation}
  K = J + a V_+ \wedge V_- + b V_+ \wedge V_3 + c V_- \wedge V_3  \; ,
\end{equation}
where $a$, $b$ and $c$ are coefficients which should be derived from
imposing that $J$ satisfies $V_\pm \lrcorner J = V_3 \lrcorner J = 0$.
From the Fierz relations with one free index we see that $V_+$ is
already orthogonal to $K$ and therefore $a=b=0$. In order to find the
coefficient $c$ we contract with $V_-$ and use the Fierz relations
with one free index and obtain
\begin{equation}
  \label{Kpar}
  K = J + \frac1{1-\alpha} V_- \wedge V_3  \; .
\end{equation}
The other cases work in a similar way. However, when considering forms of higher rank, the number of terms that can be written on the
right hand side increases rendering the calculation rather tedious.  We find, eventually, the following equations
\begin{align}
  \label{Psipar}
  \Psi & =  \varphi + \frac1{1+\alpha} J \wedge V_+
  + \frac1{2(1-\alpha)} V_+ \wedge V_- \wedge V_3 \; , \\
  \label{Phi+par}
  \Phi_+ & =  - \frac2{1+ \alpha} J \wedge J - \frac2{1+ \alpha} \rho
  \wedge V_+ - \frac2{1- \alpha} J \wedge V_- \wedge
  V_3 \; , \\
  \label{Phi-par}
  \Phi_- & =  \frac4{1 - \alpha} \varphi \wedge V_3 + \frac2{1- \alpha}
  \rho \wedge V_- + \frac2{1 + \alpha} J \wedge V_+ \wedge V_3 \; , \\
   \label{Phi3par}
  \Phi_3 & =  - \frac1{1-\alpha} \varphi\wedge V_- +
  \frac2{1-\alpha} \rho \wedge V_3 - \frac1{2(1+\alpha)}
  J \wedge V_+ \wedge V_- \; .
\end{align}
In the above the three-form $\rho$ is not independent, but can be expressed
as
\begin{equation}
  \label{rhodef}
  \rho_{mnp} = J_{rm} \varphi _{np}{}^r \; .
\end{equation}
Using the Fierz relation which involves $K$ and $\Psi$ one
can check that the RHS of the above equation is indeed antisymmetric
in all three indices as should be for the components of a 3-form (see eq. \eqref{antirho}). 
Note also that $\rho$ and $\varphi$ are orthogonal to the vectors $V_i$.

Using the Fierz relations in appendix \ref{sec:frz} it is not very
hard to check the parametrisation above. The terms which contain at
least one vector field can be immediately verified by projecting the
entire relation on the corresponding vector and using the fact that
vectors $V_\pm$ and $V_3$ are orthogonal. The only remaining problem
is to determine the top forms which are orthogonal to the
vectors. There are two different cases above. In \eqref{Psipar}, we
denote this top form by $\varphi$ and we shall further discuss its
properties. In the remaining relations, the top form is no longer an
independent form, but is given in terms of $J$, as in \eqref{Phi+par},
or it simply vanishes as in \eqref{Phi-par} and \eqref{Phi3par}. The
way to decide whether or not one can write additional terms in
\eqref{Phi+par}--\eqref{Phi3par} is by computing the norms of the RHS
and LHS of these relations. It is not difficult to check that these
norms precisely agree, and therefore if we were to add some arbitrary
form to these relations, such forms would automatically have zero norm
and thus vanish on an Euclidean space.

Naively the equations \eqref{Kpar}--\eqref{Phi3par} diverge at the
special points $\alpha= \pm1$.
To show that actually there are no divergences in the $SU(3)$
parameterisation in the $SU(4)$ or $G_2$ loci limits, we can extract
the leading behaviour with respect to $(1 \pm \alpha)$ of the relevant
$SU(3)$-structure forms
\bea
V_3 &\sim& \left(1-\alpha\right)^{\frac12} \;,\;\;\; V_- \sim \left(1-\alpha\right)^{\frac12} \;,\;\;\; V_+ \sim \left(1+\alpha\right)^{\frac12} \;, \nonumber \\
\rho &\sim& \left[\left(1-\alpha\right)\left(1+\alpha\right)\right]^{\frac12} \;,\;\;\;\; \varphi \sim \left(1-\alpha\right)^{\frac12} \;,\;\;\;\;\;\; J \sim \left(1+\alpha\right)^{\frac12} \;.
\label{alplimits}
\eea
These relations can be inferred from the norms of these objects
which we compute in appendix. The objects that define the geometry
are the 9-dimensional bilinears. These are perfectly smooth over the
full range of $\alpha$, though they take particular, different, forms
over the $SU(3)$, $SU(4)$ and $G_2$ loci.

Note that apart from $J$, $\varphi$ and $\rho$, which satisfy
\eqref{rhodef}, no other form orthogonal to the vectors appears. We
can in principle use in the parametrisation only $\varphi$ and $J$,
but it is more intuitive to consider also $\rho$ as these forms are used in
general to describe a SU(3) structure.
Indeed, the results obtained agree precisely with what is
expected from a $SU(3)$-structure in six dimensions plus three
additional directions orthogonal to it. However, in order to establish
the $SU(3)$-structure behind we still have to check a few more
relations which the forms $J$ and $\varphi$ satisfy. Using the
symmetric relation for $K$ given in eq. \eqref{symK} of the appendix
we can compute the similar identity for $J$
\begin{align}
  J_{mn} J^n{}_p & =  -\frac12 (1+ \alpha) \delta_{mp} + \frac14 (V_+)_m
  (V_+)_p + \frac{1+\alpha}{4(1-\alpha)}(V_-)_m (V_-)_p +
  \frac{1+\alpha}{1-\alpha}(V_3)_m (V_3)_p \nonumber \\
   & =  \frac12 (1+ \alpha) \left[ - \delta_{mp} + (P_+)_{mp} +
     (P_-)_{mp} + (P_3)_{mp} \right] \; ,
\end{align}
where $P_{\pm,3}$ denote the projectors on the directions $+$, $-$ and
$3$ respectively. It follows that by an appropriate normalisation $J$
can be indeed viewed as the almost complex structure on the 6-dimensional
subspace orthogonal to $V_\pm$ and $V_3$.

Making use of the Fierz identities listed in the appendix and of the eqs. \eqref{Kpar}--\eqref{Phi3par} one can show that the following identities must hold
\begin{align}
  \label{su3rel}
  J \lrcorner \varphi & =  J \lrcorner \rho = 0 \; , \\
  J \wedge \varphi & =  J \wedge \rho = 0 \; , \\
  \varphi \wedge \rho & =  *(V_+ \wedge V_- \wedge V_3) \; , \\
  \label{su3vol}
  J \wedge J \wedge J & =  \frac{3(1+ \alpha)}{2(1- \alpha)} * (V_+ \wedge
  V_- \wedge V_3) \; .
\end{align}
Furthermore, it is possible to show that eqs. \eqref{Kpar}--\eqref{su3vol} together with eq. \eqref{phiphi} imply all the Fierz identities for the bilinear forms listed in the appendix. By construction, the converse is also true.

It is useful to construct from $\varphi$ and $\rho$ the following holomorphic and anti-holomorphic (with respect $J$) three-forms
\begin{equation}
  \label{omegadef}
  \Omega = \varphi + i \sqrt{\tfrac2{1+ \alpha}} \rho \; , \qquad
  \bar \Omega = \varphi - i \sqrt{\tfrac2{1+ \alpha}} \rho \; .
\end{equation}
Indeed, it is easy to see that these forms obey the relations
\begin{equation}
  J_m{}^n \Omega_{npq} = i \sqrt{\tfrac{1+\alpha}2} \Omega_{mpq} \; , \qquad
  J_m{}^n \bar \Omega_{npq} = - i \sqrt{\tfrac{1+\alpha}2} \bar
  \Omega_{mpq} \; ,
\end{equation}
and therefore, up to some normalisation, $\Omega$ can be seen as
a $(3,0)$ form with respect to the almost complex structure $J$.
It should now be clear that following a suitable normalisation the
forms $J$ and $\Omega$ (or its real components $\varphi$ and $\rho$)
can be seen as the forms defining
an $SU(3)$-structure on the space orthogonal to the vectors $V_\pm$ and
$V_3$. Note that the normalisation we have depends on the
parameter $\alpha$ and care must be taken over the $\alpha=\pm 1$ points where a reparameterisation in terms of $SU(4)$ or $G_2$ structures is more suitable.

\section{Supersymmetry conditions}
\label{sec:susycond}

The G-structure technology introduced in the previous section is ideal for studying the supersymmetry equations (\ref{susyvar}). In this section we rewrite the supersymmetry constraints as differential constraints on the appropriate forms and use these to extract general properties of any solutions, in particular with respect to supporting a background with varying $\alpha$.

The supersymmetry variation of the gravitino \eqref{susyvar} splits
into internal and external components which read
\cite{Martelli:2003ki,Lazaroiu:2012du,Lazaroiu:2012fw}
%
\begin{equation}
  \label{susyeq}
  D_\alpha \xi = \nabla_{\alpha} \xi + A_\alpha \xi = 0 \; , \qquad Q \xi = 0 \; ,
\end{equation}
where we defined
\begin{align}
  \label{AQdef}
  A_\alpha = \lambda \gamma_\alpha \gamma_9 + \frac1{24} F_{\alpha \beta \gamma \delta} \gamma^{\beta \gamma \delta} +
  \frac14 \tilde{f}_\beta \gamma_\alpha{}^\beta \gamma_9 \; , \\
  Q = - \lambda \gamma_9 + \frac12 \partial_\alpha \Delta \gamma^\alpha -
  \frac1{288} F_{\alpha \beta \gamma \delta} \gamma^{\alpha \beta \gamma \delta} - \frac16 \tilde{f}_\alpha \gamma^\alpha \gamma_9
  \; ,
\end{align}
and the covariant derivative $\nabla_\alpha$ is taken now with respect to the 8-dimensional metric defined in \eqref{metric}.
The parameter $\lambda$ is the cosmological constant for the 3-dimensional external space.
These equations are valid for spinors $\xi$ which live on an 8-dimensional manifold. As discussed previously we are interested in studying the
background on a 9-dimensional manifold $Y_9 = X_8 \times S_1$. To do this we uplift on a circle by adding a ninth direction so that now we have an index range $m=1,...,9$. We uplift the gamma matrices by considering $\gamma_9$ together with the other gamma
matrices $\gamma_\alpha$, as the generators of the Clifford algebra $Cl(9,0)$. The analysis of the geometry and G-structures performed above does not identify a particular direction and so we would like to rewrite the uplifted supersymmetry equations also in an $SO(9)$ covariant way. We therefore introduce a constant vector field $\theta$ such that
\begin{equation}
  \gamma_9 = \theta^m \gamma_m \; , \quad m = 1, \ldots,9 \; .
\end{equation}
With this the supersymmetry equations \eqref{susyeq} have the same form where now
\begin{align}
  A_m = \lambda \theta^n \gamma_{mn} + \frac1{24} F_{mnpq} \gamma^{npq} +
  \frac14 \tilde{f}_n \theta_p \gamma_m{}^{np} \; , \label{A9d}\\
  Q = - \lambda \theta^m \gamma_m + \frac12 \partial_n \Delta \gamma^n -
  \frac1{288} F_{mnpq} \gamma^{mnpq} - \frac16 \tilde{f}_m \theta_n
  \gamma^{mn} \; , \label{Q9d}
\end{align}
and we must impose the independence of physics quantities on the ninth direction
\begin{equation}
  \label{tof}
  \theta \cdot \tilde{f} = 0 \; , \qquad \theta \lrcorner F = 0 \; , \qquad \theta \cdot d \Delta = 0 \;.
\end{equation}
These equations now hold after an arbitrary $SO(9)$ rotation which no longer identifies $\theta$ with the index value $m=9$.

Note the following simple consequences of the above. Firstly, the restrictions
on the fluxes imply that one component of $A$, i.e. $\theta^m A_m$, identically vanishes. This means
that also the spinor $\xi$ does not depend on this direction. Secondly, since we chose the 8-dimensional gamma matrices
to be real and symmetric (this also holds for $\gamma_9$), $A_m$ is a
totally antisymmetric matrix. Contracting \eqref{susyeq} with $\xi^T$
from the left we find
\begin{equation}
  0 = \xi^T \nabla_m \xi = \tfrac12 \nabla_m (\xi^T \xi) =
  \tfrac12 \partial_m ||\xi||^2\; ,
\end{equation}
and therefore the supersymmetry equations impose that the internal
spinors must have constant norms as we already anticipated in the
previous section.

The auxiliary uplift direction should be given by a linear combination of the 9-dimensional vectors. This can be seen from the analysis of the various
particular cases. In particular for the $SU(4)$ case, in 8-dimensions
we expect no singlet vector field and so the vector field which survives
should be interpreted as the additional ninth direction.
In the $G_2$ case in 8 dimensions, we expect a single vector field
which is a singlet under the structure group. In 9 dimensions we found
two such vector fields and therefore a certain linear combination of
them should be interpreted as the additional direction. Finally, in
the case of a $SU(3)$ structure we expect two singlet vector fields in
eight dimensions, while we found three of them in 9
dimensions. Again, one combination of them should precisely give the
additional direction. Thus we can write
\begin{equation}
  \label{thetadec}
    \theta = \tfrac1{2(1+ \alpha)} (\theta \cdot V_+) V_+ +
  \tfrac1{2(1- \alpha)} (\theta \cdot V_-) V_-  +  \tfrac2{1-
    \alpha} (\theta \cdot V_3) V_3 \; .
\end{equation}

Finally it is worth noting the practical matter that uplifting to 9-dimensions does not add any further complexity to the equations. This is because we are working with Majorana, rather than Majorana-Weyl, spinors in 8 dimensions which implies that they do not have any nice properties under $\gamma_9$. Therefore including it in a higher dimensional Clifford algebra is useful, in particular because the complete set of gamma matrices means the Hodge star acts simply allowing us to write all the bilinears in terms of forms of degree 4 or less.

\subsection{Equations for $N=2$ supersymmetry in 3D}
\label{sec:susyeq}

For $N=2$ supersymmetry in 3 dimensions, the supersymmetry equations
\eqref{susyeq} have to be
satisfied by two spinors on the internal manifold, $\xi_{1,2}$.
In order to see what sort of constraints these equations impose on the
9-dimensional manifolds discussed in the previous section we should first rewrite
them in terms of the spinor bilinears \eqref{bilinears}. Using the
definitions \eqref{bilinears} and \eqref{susyeq} we can compute the
derivatives of all the spinor bilinears to find \cite{Lazaroiu:2012fw}
\begin{align}
  \label{dforms}
  \nabla_m V_{i\; n} & =  2 \lambda \theta_n V_{i \; m} - 2 \lambda
  (\theta \cdot V_i) \delta_{mn} - \frac1{12} F_{mpqr} \Phi_{i\;
    n}{}^{pqr} + \frac12 \Phi_{i \; mnpq} \tilde{f}^p \theta^q \\
  \nabla_m K_{np} & =  -4 \lambda K_{m [n} \theta_{p]} + 4 \delta_{m[n}
  K_{p]q} \theta^q + \frac12 F_{m[n}{}^{qr} \Psi_{p]qr} +
  \Psi_{m[n}{}^q \theta_{p]} \tilde{f}_q \nonumber \\
  &  -  \Psi_{m[n}{}^q \tilde{f}_{p]} \theta_q + \delta_{m[n} \Psi_{p]qr} \tilde{f}^q
  \theta^r \\
  \nabla_m \Psi_{npr} & =  6 \lambda \Psi_{m[np} \theta_{r]} - 6
  \lambda \delta_{m[n} \Psi_{pr]q} \theta^q + \frac1{12} F_m{}^{stu}
   (*\Psi)_{nprstu} \nonumber \\
  &  + \frac32 F_{m[np}{}^q K_{r]q} - \frac1{2}
  (*\Psi)_{mnprst} \tilde{f}^s \theta^t - 3 K_{m[n} \tilde{f}_p \theta_{r]} \\
  &  + 3 \delta_{m[n} \tilde{f}_p K_{r]q} \theta^q - 3 \delta_{m[n} \theta_p
  K_{r]q} \tilde{f}^q \; , \nonumber \\
  \nabla_m \Phi_{i \; npqr} & =  - 8 \lambda \Phi_{i \; m[npq}
  \theta_{r]} + 8 \lambda \delta_{m[n} \Phi_{i \; pqr]s} \theta^s +
  F_{m[n}{}^{st} (* \Phi_i)_{pqr]st} \nonumber \\
  &  - 2 F_{m[npq} V_{i \; r]} + 2 (* \Phi_i)_{m[npq}{}^s \theta_{r]}
  \tilde{f}_s - 2 (*\Phi_i)_{m[npq}{}^s \tilde{f}_{r]} \theta_s \\
  &  + 2 \delta_{m[n} (*\Phi_i)_{pqr]st} \tilde{f}^s \theta^t - 12
  \delta_{m[n} V_{i \; p} \tilde{f}_q \theta_{r]} \; , \nonumber
\end{align}
where the subscript $i$ stands for any of $1, 2, 3, +$ or $-$.
It is interesting to note that the equations above imply that the
derivatives along $\theta$ vanish identically upon using the condition
on the fluxes \eqref{tof}.

We should now consider the constraints which come from the external
gravitino variation. These equations should be projected on a basis of
spinors in order to find an equivalent set of equations. In 8
dimensions it is easy to find a
basis of spinors in terms of Majorana-Weyl singlet spinors, but since
the Majorana-Weyl components of the spinors we consider may vanish at
certain points, we will not be able to use a basis constructed in such
a way globally. Instead we shall
project the supersymmetry equations on a larger set of spinors -- not linearly
independent -- which are constructed by multiplying the spinors
$\xi_{1,2}$ by arbitrary elements of the Clifford
algebra. Specifically we shall project the spinor equations
\eqref{susyeq} on spinors of the form
\begin{equation}
  \xi_{i \; A} = \gamma_A \xi_i \; , \quad i = 1,2\; , \qquad \gamma_A
  \in Cl(9,0)
\end{equation}
In practice we shall see that taking $\gamma_A = \mathbbm{1}, ~
\gamma_m$ is enough and the other constraints are just consequences of
these ones. Therefore we shall use the following equations
\begin{eqnarray}
  \label{Qcon01}
  - \lambda (\theta \cdot V_i) + \frac12 (V_i \cdot d \Delta) -
  \frac1{12} F \lrcorner \Phi_i = 0 \\
  \label{Qcon02}
  \tilde{f}_m \theta_n K^{mn} = 0 \\
  \label{Qcon11}
  - 2 \lambda \theta + d \Delta - \frac1{12} *(F \wedge \Phi_+)
  + \frac16 V_+ \lrcorner (\tilde{f} \wedge \theta) =  0 \; \\
  \label{Qcon12}
  - \frac1{12} *(F \wedge \Phi_{-\;,3}) + \frac16 V_{- \; , 3}
  \lrcorner (\tilde{f} \wedge \theta) =  0 \; \\
  \label{Qcon13}
  2 \lambda \theta \lrcorner K - d \Delta \lrcorner K + \frac16 \Psi
  \lrcorner F - \frac13(\tilde{f} \wedge \theta) \lrcorner \Psi =  0
\end{eqnarray}
while the ones corresponding to other Clifford elements can be found in
appendix \ref{sec:susyeqapp}.

\subsection{Flux induced variations of $\alpha$}
\label{sec:varstru}

The key property of the backgrounds that we are interested in is the variation of the parameter $\alpha$, since this is the property that signals an M2-brane potential. We will explain this relationship in more detail in section \ref{sec:m2pot}. In this section we are interested in determining which flux is responsible for inducing such a variation. In general not any flux will induce this variation, for example we know that $(2,2)$ primitive flux as studied in the case of $SU(4)$-structure compactifications \cite{Becker:1996gj} will not induce such a variation. We will determine the flux which is relevant by using the supersymmetry equations to directly evaluate the variation of $\alpha$.

The form of the supersymmetry equations as given in section \ref{sec:susyeq} carries redundancies because
the bilinears \eqref{bilinears} are not independent, but satisfy various relations coming from the Fierz
relations. In order to find a more tractable set of equations we should use a parametrisation for the forms \eqref{bilinears} which
already makes use of the Fierz relations. A particular parameterisation is the one given in \eqref{Kpar}--\eqref{Phi3par} which is
valid for the $SU(3)$-structure loci. We claim that this parameterisation is sufficient to capture all the variation of $\alpha$. This follows from the simple
reasoning that over the patches where it breaks down, the $SU(4)$ and $G_2$ loci, $\alpha$ is constant by definition. Further it is also at a maximum or minimum value over these loci so that derivatives along directions leading away from the loci also vanish. Therefore over such loci $d \alpha =0$ and all the variation is within the $SU(3)$-structure locus.

Using the supersymmetry equations (\ref{dforms}), the variation of $\alpha$ can be generally computed as
\begin{equation}
  \label{dalpha0}
  \nabla_m \alpha = \frac12 \nabla (V_+^n V_{+\; n})= V_+^n \nabla_m V_{+ \; n} = - \frac1{12}V_+^n F_{mpqr} \Phi_{+\; n}{}^{pqr} +
  \frac12 \Phi_{+\; mnpq} \tilde{f}^p \theta^q V_+^n \; .
\end{equation}
Making use of the $SU(3)$ parametrisation of the form $\Phi_+$,
\eqref{Phi+par}, and of the fact that the forms $J$, $\rho$, $V_-$ and
$V_3$ are all orthogonal to $V_+$, together with the fact that
$\theta$ should be a combination of the vector fields $V_\pm$ and
$V_3$, we find that the second term above does not contribute to the
variation of the parameter $\alpha$ and so
\begin{equation}
  \label{dalpha}
  \nabla_m \alpha = - \frac13 \rho_{pqr} F_m{}^{pqr} \; .
\end{equation}
Given again all the orthogonality properties of the $SU(3)$-structure
forms we find that the flux which is responsible for the variation of
$\alpha$ can be written as
\begin{equation}
  \label{niceflux}
  F = \tilde h \wedge \rho + \tilde g \wedge \varphi \; ,
\end{equation}
where at this stage $\tilde h$ and $\tilde g$ are arbitrary one-forms
on the internal manifold that parameterise the flux.

Note that $\rho$ and $\varphi$ are (depending on the conventions) the real and imaginary part of the
holomorphic 3-form which can be defined on a manifold with $SU(3)$-structure.
Over the $SU(4)$ locus the flux therefore lifts to either $(4,0)$+$(0,4)$ or $(3,1)$+$(1,3)$ flux.
Note that both these fluxes are forbidden by supersymmetry in the case of pure $SU(4)$-structure where only (2,2)
primitive fluxes are allowed \cite{Becker:1996gj}, which is consistent with the understanding that the variation of $\alpha$ vanishes over that locus.

\subsection{M2-brane potentials and Supersymmetry}
\label{sec:m2pot}

At this stage it is worth going into some more detail regarding the relation between a potential for probe space-time filling M2-branes and the background geometry. In type IIB compactified on manifolds with $SU(3) \times SU(3)$ structure, we know that
space-time-filling D3 branes become supersymmetric at points where the manifold locally looks like
a manifold with SU(3) structure (i.e. the two spinors defining
the $SU(3) \times SU(3)$ structure are parallel) \cite{Martucci:2006ij,Koerber:2006hh}. In this section we will find similar results for M2-branes in terms of local structure groups.

Let us briefly recall some well-known facts about supersymmetry and
M2 branes. We will utilise the form of the M2-brane action given in \cite{Marolf:2003vf}
\begin{equation}
  S_{M2} = - T_{M2} \int d^3 \zeta \sqrt{-G} - \frac{T_{M2}}{6} \int d^3
  \zeta \epsilon^{ijk} A_{kji} + i \frac{T_{M2}}{2} \int d^3\zeta
  \sqrt{-G} \bar y (1 - \Gamma_{M2}) \Gamma^i \tilde D_i y \;, \label{m2act}
\end{equation}
where $G_{ij}$ is the induced metric on the brane, $T_{M_2}$ is the
tension of the brane, $A_{ijk}$ is the pull-back of the supergravity
3-form on the brane, $\tilde D_i$ is the pull-back of the
supercovariant derivative and $y$ is an 11-dimensional Majorana spinor. $\Gamma_{M2}$ is the brane chirality
operator which is given by
\begin{equation}
  \Gamma_{M2} = \frac{1}{6 \sqrt{-G}} \epsilon^{ijk} \Gamma_i \Gamma_j \Gamma_k \;.
\end{equation}
Note that $\Gamma_{M2}^2 = 1$ so that its eigenvalues are $\pm 1$ and
we can define projectors on the corresponding subspaces as
\begin{equation}
  P_{\pm} = \frac12 (1 \pm \Gamma_{M2}) \; .
\end{equation}
The action (\ref{m2act}) is invariant under local $\kappa$ transformations \cite{Marolf:2003vf}
\begin{equation}
  \delta_\kappa y = (1 + \Gamma_{M2}) \kappa + \mathcal{O}(y^2) \; ,
  \qquad \delta_{\kappa} X^M = \frac{i}{2} \bar y \Gamma^M (1 +
  \Gamma_{M2}) \kappa + \mathcal{O}(y^3)
\end{equation}
where $\kappa$ is a 32-component spinor which may depend on the
coordinates on the worldvolume of the M2 brane, $\zeta^i$.

For a background with Killing spinor $\epsilon$ satisfying (\ref{susyvar}), supersymmetry transformations act as
\begin{equation}
  \delta_{\epsilon} y = \epsilon + \mathcal{O}(y^2) \; , \qquad
  \delta_{\epsilon} X^M = - \frac{i}{2} \bar y \Gamma^M \epsilon +
  \mathcal{O} (y^3) \; .
\end{equation}

A bosonic brane configuration ($y=0$) is supersymmetric only if
$\delta_{\epsilon} y = 0 $ and we see that this can not be satisfied
unless $\epsilon=0$. However $\delta_{\epsilon} y = \epsilon$ is
compatible with supersymmetry only if this transformation can be
compensated by a $\kappa$ transformation \cite{Becker:1995kb}. Therefore the
only supersymmetry generators which are not broken by the bosonic
brane configuration are those which can be written as
\begin{equation}
  \epsilon= \delta_{\kappa} y = (1 + \Gamma_{M2}) \kappa = 2P_{+}
  \kappa \; .
\end{equation}
This is equivalent to requiring \cite{Becker:1995kb}
\begin{equation}
  \label{susyspinor}
  P_{-} \epsilon = (1- \Gamma_{M2} ) \epsilon = 0  \; .
\end{equation}
For the compactification to 3 dimensions with space-time-filling
M2-branes, the M2 brane chirality operator decomposes according to the
the split of gamma matrices \eqref{cliff}, as
\begin{equation}
  \Gamma_{M2} = \mathbf{1} \otimes \gamma_9
\end{equation}
where $\gamma_9$ is the chirality operator on the 8d manifold.
Using the spinor decomposition for the compactification to 3
dimensions \eqref{spdec} we find we find that the condition above is
equivalent to
\begin{equation}
  \label{M2susy}
  \gamma_9 \xi =\xi  \; .
\end{equation}
We see that the condition that the M2 brane is supersymmetric requires Killing spinors of definite chirality.

With this result in mind consider probe M2 branes in our supergravity
background at the different local $SU(3)$, $SU(4)$ or $G_2$-structure
loci. Generally we have that there are 2 Majorana Killing
spinors. Over $G_2$ loci, $\alpha=-1$, these become Majorana-Weyl
Killing spinors of opposite chirality. Therefore M2 branes preserve
$N=1$ supersymmetry on these loci. Over $SU(4)$ loci, $\alpha=+1$, the
Killing spinors are Majorana-Weyl of same chirality and so M2 branes either preserve $N=2$ supersymmetry or no supersymmetry. However, for a given
fixed chirality of the two Majorana-Weyl spinors either M2 or anti-M2 branes preserve $N=2$ supersymmetry, and we define M2 branes as the objects which
preserve the supersymmetries over $SU(4)$ loci.\footnote{Note that there exists the interesting possibility of multiple $SU(4)$ loci with different chiralities in which case neither M2 or anti-M2 can preserve supersymmetry on all of loci. We will not consider such configurations in detail in this work.} Over the $SU(3)$ loci, $-1< \alpha < 1$, we have two Killling Majorana spinors which contain 4 non-vanishing
Majorana-Weyl component spinors. We have three possibilities: either all
of the Majorana-Weyl components are Killing individually, in which case the
M2 branes preserves $N=2$ supersymmetry, or two Majorana-Weyl components of one Majorana spinor are Killing while the components of the other are not, in which case the M2 preserves $N=1$ supersymmetry, or none of the components are
Killing in which case an M2 brane is non-supersymmetric. In appendix
\ref{alflux} we show that the flux which induces a varying $\alpha$
(\ref{niceflux}) precisely implies that the last possibility is
realised and M2 branes preserve no supersymmetry. Indeed it is rather
simple to see that this should be the case for any background which
does not have a global $SU(3)$ or $G_2$-structure since if any Majorana-Weyl components were covariantly constant by themselves their
norm could be set to unity over the full manifold, thereby implying
that they are nowhere vanishing and induce a global $SU(3)$ or $G_2$-structure.

On general grounds we expect that supersymmetric loci are minima of the potential. This means that on the generic (non-supersymmetric) locus a probe M2 brane should feel a potential which drives it to a supersymmetric locus.

Let us be more explicit about the condition for $N=2$ supersymmetry of
the M2 brane. The condition \eqref{M2susy} has to be satisfied for two
spinors $\xi_1$ and $\xi_2$, which means that the two spinors are
actually Majorana--Weyl of positive chirality and therefore define a
SU(4) structure. In the language used in section \ref{sec:susy8d} this means
that the vectors $V_1$ and $V_2$ are equal and therefore
\begin{equation}
  \label{M2sol}
  V_- =0 \; .
\end{equation}
Conversely, the condition $V_-=0$ implies that we are dealing with a
SU(4) point and therefore the two spinors are
Majorana--Weyl.\footnote{In principle the chirality of the spinors can
not be determined. However, if this is negative, and therefore, the
condition \eqref{M2susy} is not satisfied, this would be a point where
anti M2 branes are supersymmetric.} Finally, it is interesting to note that, in the spirit of the analysis in \cite{Martucci:2006ij,Koerber:2006hh} of D3 superpotentials, the condition \eqref{M2sol} hints that the 1-form $V_-$ may be associated to the derivative of a world-volume potential.

\section{Analysis of the special flux}
\label{sec:anapartfl}

In the previous section we identified the particular flux that sources the variation of $\alpha$. In this section we study backgrounds that can support this flux. As an initial investigation we restrict the 4-form flux $F$ to be solely composed of the flux of interest so that it takes the form (\ref{niceflux}), while the warp-factor $\Delta$ and 1-form flux $f$ are unconstrained. We leave a complete investigation allowing also for other types of 4-form fluxes in the background for future work.

It is useful to decompose the 4-form flux along the three vectors as
\begin{equation}
  \label{nicefluxdec}
  \begin{aligned}
    F =~& h \wedge \rho + g \wedge \varphi \\
    & + \tfrac1{2(1+\alpha)} h_+ V_+ \wedge \rho + \tfrac1{2(1-\alpha)}
    h_- V_- \wedge \rho + \tfrac2{(1-\alpha)} h_3 V_3 \wedge \rho \\
     & + \tfrac1{2(1+\alpha)} g_+ V_+ \wedge \varphi + \tfrac1{2(1-\alpha)}
    g_- V_- \wedge \varphi + \tfrac2{(1-\alpha)} g_3 V_3 \wedge \varphi \;.
  \end{aligned}
\end{equation}
In the above we defined
\begin{equation}
  \label{ghcomp}
  h_i = V_i \cdot \tilde h \equiv V_i^m \tilde h_m \; , \qquad  g_i
  = V_i \cdot \tilde g \equiv V_i^m \tilde g_m \; , ~~ i = \pm, 3 \;,
\end{equation}
while $g$ and $h$ are defined as the components of $\tilde g$ and $\tilde
h$ orthogonal to the vectors $V_\pm$ and $V_3$.
We therefore decompose the fluxes $h$, $g$ and $\tilde{f}$ as
\begin{align}
  \tilde h & =  h + \frac1{2(1+\alpha)} h_+ V_+  + \frac1{2(1-\alpha)}
  h_- V_-  + \frac2{(1-\alpha)} h_3 V_3  \;, \\
  \tilde g & =  g + \frac1{2(1+\alpha)} g_+ V_+  + \frac1{2(1-\alpha)}
  g_- V_-  + \frac2{(1-\alpha)} g_3 V_3 \;,\\
  \tilde f & =  f + \frac1{2(1+\alpha)} f_+ V_+  + \frac1{2(1-\alpha)}
  f_- V_-  + \frac2{(1-\alpha)} f_3 V_3 \;,
\end{align}
where $f$, $f_\pm$ and $f_3$ are defined in analogy with $g$ and $h$ by their relation to the vectors.
Note that there are no divergences at $\alpha = \pm 1$ in the expressions above, which can be seen using the limits (\ref{alplimits}).
Recall that we are still formally working on a nine-dimensional
manifold and in order not to alter the physics we have to impose
\eqref{tof} on the fluxes above. Since $\theta$ is a linear
combination of the vectors, this condition imposes the orthogonality
of the one-form fluxes $\tilde h$, $\tilde g$ and $\tilde f$ on $\theta$
\begin{equation}
  \label{tofgh}
  \theta \lrcorner \tilde h =   \theta \lrcorner \tilde g =   \theta
  \lrcorner \tilde f =  0 \;.
\end{equation}

With these definitions, and using the various relations in appendix \ref{sec:SU3app}, we find that \eqref{dalpha} yields
\begin{equation}
  \label{dalpha1}
  d \alpha = - (1- \alpha)(1+\alpha) h - 2 (1-\alpha) g \lrcorner J -
  (1-\alpha) h_+ V_+ - (1+ \alpha) h_- V_- -  4 (1+ \alpha) h_3 V_3
\end{equation}
Note that, as expected, $d \alpha = 0$ on the $SU(4)$ and $G_2$ loci.

To analyse the supersymmetry conditions for this particular flux
Ansatz we shall start with the constraints \eqref{Qcon01}-\eqref{Qcon13}. Inserting the SU(3) parametrisation \eqref{Psipar}-\eqref{Phi3par} and
the flux Ansatz above, by using the relations in appendix \ref{sec:SU3app}
we find
\begin{subequations}
  \label{Qfgh}
  \begin{align}
    \label{Qfgha}
    0 =& -\lambda (\theta \cdot V_+)+\frac{1}{2}(V_+\cdot
    d\Delta)-\frac{1}{6}(1-\alpha) h_+ \;,\\
    \label{Qfghb}
    0 =&-\lambda (\theta \cdot V_-)+\frac{1}{2}(V_-\cdot
    d\Delta)+\frac{1}{6}(1+\alpha)h_-+\frac23 g_3 \;,\\
    \label{Qfghc}
    0 =&-\lambda (\theta \cdot V_3)+\frac{1}{2}(V_3\cdot
    d\Delta)+\frac{1}{6}(1+\alpha) h_3 -\frac{1}{6} g_- \;,\\
    \label{Qfghd}
    0 =&(\theta \cdot V_3) f_- - f_3 (\theta \cdot V_- ) \;,\\
    \label{Qfghe}
    0 =&d\Delta - 2\lambda\theta-\frac{1}{3} (g_- V_3 - g_3 V_-) +
    \frac{1}{6}f_+ \theta - \frac16 ( \theta \cdot V_+) \tilde f \;,\\
    \label{Qfghf}
    0 =&-h_+ V_- + h_- V_+ - 2 g_+ V_3+ 2 g_3 V_+ + f_-\theta - (\theta
    \cdot V_-) \tilde f \;,\\
    \label{Qfghg}
    0 =&-2 h_+ V_3 + 2 h_3 V_+ + g_+ V_- - g_- V_+ + 2 f_3 \theta - 2
    (\theta \cdot V_3) \tilde f \;, \\
    \label{Qfghh}
    0 =&\frac{2\lambda}{1-\alpha}(\theta\cdot
    V_-)V_3-\frac{2\lambda}{1-\alpha}(\theta\cdot V_3)V_- - d\Delta
    \lrcorner J -\frac{1}{1-\alpha}(d\Delta \cdot
    V_-)V_3+\frac{1}{1-\alpha}(d\Delta\cdot V_3)V_- \nonumber\\
    &+\frac{1}{6}(1-\alpha) \tilde h \lrcorner J -\frac{1}{6}(1-\alpha)
    \tilde g -\frac{1}{12}\frac{1-\alpha}{1+\alpha}g_+ V_+
    -\frac{1}{3} g_3 V_3
    -\frac{1}{12} g_-V_- \nonumber \\
    &+ \frac{(\theta  \cdot V_+)}{3(1+\alpha)} \tilde f \lrcorner J -
    \frac1{6(1-\alpha)} (\tilde f \wedge \theta) \lrcorner (V_+ \wedge V_-
    \wedge V_3) \;.
  \end{align}
\end{subequations}
Let us look more carefully at equation \eqref{Qfghf}. Contracting it
with $V_-$ we immediately find that $h_+ = 0$. By contracting with
$V_3$ and using \eqref{Qfghd} we find that $g_+=0$. Furthermore,
projecting equations \eqref{Qfghf} and \eqref{Qfghg} on $V_+$ and on
the 6-dimensional space orthogonal to the vectors $V_\pm$ and $ V_3$
one obtains the following equations
\begin{align}
\label{g3}
g_3 & = -\frac{1}{2}h_- - \frac{1}{4(1+\alpha)}f_- (\theta\cdot V_+) + \frac{1}{4(1+\alpha)}(\theta\cdot V_- )f_+ \;,\\
\label{g-}
g_- & = 2 h_3 + \frac{1}{1+\alpha}f_3 (\theta\cdot V_+) - \frac{1}{1+\alpha}(\theta\cdot V_3) f_+ \;,
\end{align}
and
\begin{equation}
  \label{tvf}
  (\theta \cdot V_-) f = 0 \;,\qquad (\theta \cdot V_3) f = 0 \;.
\end{equation}
Now, using \eqref{thetadec}, the relations coming from the
orthogonality of $\theta$ on the fluxes $\tilde f$ and $\tilde g$ become
\begin{align}
h_- (\theta\cdot V_-) + 4 h_3 (\theta \cdot V_3) & = 0 \;,\\
h_3 (\theta\cdot V_-) - h_- (\theta\cdot V_3) &= 0 \; ,
\end{align}
where we used \eqref{g3}, \eqref{g-} and \eqref{Qfghd}. These
equations can be viewed as a linear system for $h_-$ and $h_3$ which has
a non-trivial solution only if the corresponding determinant vanishes.
Therefore, there are two cases to consider. One of them must have $\theta \cdot V_- = \theta \cdot
V_3 = 0$ and the other one has to satisfy $h_- = h_3 = 0$ and,
following \eqref{tvf}, $f=0$. We shall focus on the first solutions as we
want to study loci with $d\alpha \neq 0$. Indeed, it is easy to show that in the latter case
one must have that $d\alpha = 0$. In order to prove it one needs also the identity
\begin{equation}
g\lrcorner J+ \frac{1}{2}(1+\alpha)h  = 0 \; ,
\end{equation}
resulting from eq. \eqref{Qfghh} after contraction with $J$.

Let us continue with the case $\theta \cdot V_- = \theta \cdot
V_3 = 0$. Making use of eq. \eqref{thetadec} we obtain that $\theta$
has to be in the direction of $V_+$. We can therefore write
\begin{equation}
  \label{tVp}
  \theta = \frac{(\theta \cdot V_+)}{2(1+ \alpha)} V_+ \; ,
\end{equation}
and the fact that the physics should not depend on $\theta$ now
transfers to $V_+$.
Notice that this is consistent with the expectation that the
flux should vanish over $G_2$ structure loci since in this case we
have $V_+=0$ but the auxiliary direction $\theta$ must be
non-vanishing.

Together with equation \eqref{Qfgha}, \eqref{tVp} immediately implies
that $\lambda =0$, and therefore all such compactifications are to
3-dimensional Minkowski space.  Furthermore, equations \eqref{Qfghb},
\eqref{Qfghc}, \eqref{Qfghf} and \eqref{Qfghg} allow to solve for the
projections of $d \Delta$ on $V_-$ and $V_3$ in terms of $f_-$ and
$f_3$. Altogether the equations \eqref{Qfgh} become
\begin{subequations}
  \label{Ql0}
  \begin{align}
    0 & = \lambda = f_+ = g_+ = h_+ = d \Delta \cdot V_+ \;, \\
    d \Delta \cdot V_- &= \frac{1-\alpha}3 h_- + \frac{\theta \cdot
      V_+}{3(1+ \alpha)} f_-  \;, \qquad
    d \Delta \cdot V_3 = \frac{1-\alpha}3 h_3 + \frac{\theta \cdot
      V_+}{3(1+ \alpha)} f_3 \;, \\
    g_3 &= - \tfrac12 h_- - \frac{\theta \cdot V_+}{4(1+ \alpha)} f_-
    \;, \qquad
    g_- = 2 h_3 + \frac{\theta \cdot V_+}{1+ \alpha} f_3 \;, \\
    \label{Ql0d}
    0 &= d\Delta - \frac{1}{3} (g_- V_3 - g_3 V_-) - \frac16 (
      \theta \cdot V_+  ) \tilde f \;,\\
    \label{Ql0e}
    0 &=- d \Delta \lrcorner J  + \frac{1}{6}(1-\alpha)(h \lrcorner J - g)
    + \frac{(\theta \cdot V_+)}{3(1 + \alpha)} f \lrcorner J \;.
  \end{align}
\end{subequations}
Contracting \eqref{Ql0d} with $J$ we find
\begin{equation}
  d \Delta \lrcorner J = \frac{\left(\theta  \cdot V_+ \right)}6 f \lrcorner J \;,
\end{equation}
and then \eqref{Ql0e} becomes
\begin{equation}
  \label{hgf}
  g - h \lrcorner J = \frac{\left( \theta \cdot V_+\right)}{1+ \alpha} f \lrcorner J \;.
\end{equation}
%


\subsection{The case $\tilde f=0$}
\label{sec:f=0}

At this point we split the analysis to two cases, the simpler case $\tilde{f}=0$ is studied in this section, while the more general case is studied in section \ref{sec:fneq0}. For this section we therefore set $f=f_+=f_-=f_3=0$.

Let us analyse more closely equation \eqref{hgf}. Since $J$, up to
normalisation, acts like an almost complex structure on the
6-dimensional space orthogonal to the vectors $V_\pm$ and $V_3$, this
equation tell us that a particular combination of the fluxes $g$ and
$h$ is holomorphic with respect to the almost complex structure
$J$. It is not hard to see that the complex flux defined as
\begin{equation}
  \label{holhg}
  \hh = h + i \sqrt{\tfrac2{1+ \alpha}} g \; , \qquad   \bar \hh = h -
  i \sqrt{\tfrac2{1+ \alpha}} g
\end{equation}
is holomorphic in that it obeys
\begin{equation}
  J_m{}^n \hh_n = i \sqrt{\tfrac{1+\alpha}2} \hh_m \; , \qquad
  J_m{}^n \bar \hh_n = - i \sqrt{\tfrac{1+\alpha}2} \bar \hh_m \; .
\end{equation}
It is useful to use this flux combination together with objects which
again have well-defined behavior when contracted with $J$. In
particular, we can construct a $(4,0)$ form by taking the exterior
product of \eqref{omegadef} with \eqref{holhg}.
%
Since $J$ only acts on a 6-dimensional subspace, a $(4,0)$ form in the
sense defined above must identically vanish. In particular we have \eqref{kxp}
\begin{equation}
  0 = \hh \wedge \Omega = h \wedge \varphi - \frac2{1+ \alpha} g
  \wedge \rho + i \sqrt{\tfrac2{1+ \alpha}} \left(h \wedge \rho + g
    \wedge \varphi \right) = 0 \; .
\end{equation}
Note that the imaginary part is proportional to the projection of the
flux $F$ orthogonal to the vectors. The fact that this part of the
flux vanishes implies that the variation of
$\alpha$ only depends on the projection of the $\tilde{h}$ and $\tilde{g}$ fluxes
along the vectors $V_-$ and $V_3$.

The equations \eqref{Ql0} now become
\begin{equation}
      d \Delta \cdot V_- = \frac{1-\alpha}3 h_- \; , \quad
    d \Delta \cdot V_3 = \frac{1-\alpha}3 h_3 \; , \quad
    g_3 = - \tfrac12 h_- \; , \quad  g_- = 2 h_3 \; , \quad
    h \lrcorner J - g = 0
\end{equation}
with all the rest of the flux components being zero.
All the above relations greatly simplify the differential equations
which now become
\begin{align}
  \label{df0}
  d \alpha & =  - (1+ \alpha) h_- V_-  -4 (1+ \alpha) h_3 V_3 \\
  d V_+ & =  \tfrac12 h_- V_+ \wedge V_- + 2 h_3 V_+ \wedge V_3 \; ,
  \quad d V_- =2 h_3 V_- \wedge V_3 \; , \quad d V_3 = - \frac12 h_-
  V_- \wedge V_3 \; , \\
  dK & =  - h_- J \wedge V_- - 4 h_3 J \wedge V_3 \; , \\
  d \Psi & =  \frac2{1+ \alpha} J \wedge V_+ \wedge \left(\tfrac14 h_- V_- + h_3 V_3 \right) - \frac2{1 - \alpha} \rho \wedge  (h_- V_3 - h_3
  V_-) \nonumber \\
  &  + \frac{1- 5 \alpha}{1 -\alpha} \varphi \wedge \left(
  \tfrac14 h_- V_- + h_3 V_3 \right) \; ,
\end{align}
where we made extensive use of the relations \eqref{kcon} for $k=h$
which, in the case $\tilde f=0$, implies $\tilde k = g$.
Note that the exterior derivative of $J$ is the same as the derivative
of $K$ above, while for the derivative of $\varphi$ we find
\begin{equation}
  d \varphi =  \frac2{1- \alpha} \rho \wedge (h_3 V_- - h_- V_3) +
  \frac{1 - 5 \alpha}{1- \alpha} \varphi \wedge \left(\tfrac14 h_- V_- +  h_3 V_3\right)
\end{equation}
In the above formula the term which is in the direction of $V_+$ in $d
\Psi$ drops out. This is precisely as it should be, as $\varphi$ has no
components along $V_+$ and moreover its derivative along $\theta$
(which is identified with $V_+$ in this case) vanishes.
With a bit of more work one can compute the exterior derivative of
$\rho$ as well. We find
\begin{equation}
  d \rho = \frac{3- 7 \alpha}{1-\alpha} \rho \wedge \left(\tfrac14 h_- V_- + h_3 V_3
     \right) + \frac{1+ \alpha}{1- \alpha} \varphi
  \wedge \left(h_- V_3 - h_3 V_- \right)
\end{equation}
At a first glance it seems that $d \varphi$ and $d \rho$ do not
combine nicely into $d \Omega$, but one has to take into account that
$d \Omega$ contains an additional term of the form $d \alpha \wedge
\rho$ due to the $\alpha$-dependent factor in front of $\rho$ in the
definition of $\Omega$. With this we find
\begin{equation}
  d \Omega = \frac{1- 5 \alpha}{1- \alpha} \Omega \wedge \left(\tfrac14  h_- V_- + h_3 V_3
   \right) + i \frac{1 + \alpha}{1- \alpha} \sqrt{\frac2{1
      + \alpha}} \Omega \wedge (h_- V_3 - h_3 V_-) \;.
\end{equation}
It is important to notice that the exterior derivatives of the $SU(3)$-structure
forms do not have components strictly orthogonal to the
vectors. Therefore one can conclude that the intrinsic torsion classes
for the 6-dimensional manifold orthogonal to the vectors vanish and therefore
this is a Calabi--Yau manifold. We conclude that in the case $\tilde
f=0$ the supersymmetry equations require that, over the $SU(3)$ locus, the 8-dimensional manifold is a
fibration of a 6-dimensional CY manifold over a 2-dimensional base spanned by the vectors
$V_-$ and $V_3$. While over any $G_2$ or $SU(4)$ loci the flux vanishes.

Finally let us note that using the normalisation of $V_+$ we can write
\begin{equation}
  V_+ = \sqrt{2(1+ \alpha)} \theta \;,
\end{equation}
taking $\theta$ to be constant we derive
\begin{equation}
  d V_+ = \frac1{2(1+ \alpha)} d \alpha \wedge V_+ \; ,
\end{equation}
which precisely agree with $d V_+$ found in \eqref{df0}. This is
consistent with the interpretation that the auxiliary 9-dimensional manifold is a
direct product of the original 8-dimensional manifold and a circle.

\subsection{The case $\tilde f \ne 0$}
\label{sec:fneq0}

Let us now return to the study of the $\tilde f \ne 0$ case. Compared
to the case $\tilde f = 0$ we see that now the complex fluxes
\eqref{holhg} are no longer (anti)holomorphic. Rather one should
replace $h \to h + \frac{( \theta \cdot V_+)}{1+ \alpha} f$ in order to
obtain a holomorphic combination. This means that following the same argument of constructing a $(4,0)$ form on the space orthogonal to the vectors introduced in the previous section, we now find
\begin{equation}
  h \wedge \rho + g \wedge \varphi = - \frac{\left(\theta  \cdot V_+\right)}{1+
    \alpha} f \wedge \rho \; ,
\end{equation}
and so the flux $F$ has also a component which is orthogonal to the
vectors.

For the variation of $\alpha$ we find
\begin{equation}
  d \alpha = (1-\alpha) (\theta \cdot V_+) f - (1+ \alpha)( h_- V_- + 4 h_3
    V_3) \; .
\end{equation}
One can compute again the derivatives of the forms and we find
\begin{align}
  d V_+ & =  \frac{1- \alpha}{2(1+ \alpha)} (\theta \cdot V_+) f \wedge V_+ + 2
  h_3 V_+ \wedge V_3 + \frac12 h_- V_+ \wedge V_- \;, \\
  d V_- & =  \frac{2 (\theta \cdot V_+)}{(1- \alpha)(1 + \alpha)} f_3
  V_- \wedge V_3 + 2 h_3 V_- \wedge V_3 - \frac12(\theta \cdot
    V_+) f \wedge V_- \;,\\
  d V_3 & =  -\frac{(\theta \cdot V_+)}{2 (1- \alpha)(1 + \alpha)} f_-
  V_- \wedge V_3 - \frac12 h_- V_- \wedge V_3 - \frac12 (\theta \cdot
    V_+) f \wedge V_3 \;,\\
  dK & = -h_- J \wedge V_- -4 h_3 J \wedge V_3 - \frac{\alpha}{1+\alpha}(\theta \cdot V_+) f \wedge J - \frac{\theta\cdot V_+}{2(1-\alpha)}f\wedge V_-\wedge V_3 \nonumber\\
  & - \frac{\theta\cdot V_+}{2(1+\alpha)(1-\alpha)}f_- J\wedge V_- - 2\frac{\theta\cdot V_+}{(1+\alpha)(1-\alpha)}f_3 J\wedge V_3 \;.
\end{align}
$d\Psi$ can again be computed directly from its covariant
derivative. Again, when deriving $d \varphi$ the terms in the
direction of $V_+$ cancel and we are left with
\begin{align}
  d \varphi & =  \frac32 (\theta \cdot V_+) \varphi \wedge f +
  \frac{(\theta \cdot V_+)}{2(1- \alpha)} (f \lrcorner \rho) \wedge
  V_- \wedge V_3 + \frac{1- 5 \alpha}{1 - \alpha} \varphi \wedge \left( \tfrac14 h_- V_- + h_3
  V_3  \right ) \nonumber \\
  &  + \frac{3 (\theta \cdot V_+)}{(1-\alpha)(1+ \alpha)} \varphi
  \wedge \left(\tfrac14 f_- V_- + f_3 V_3  \right) - \frac2{1- \alpha} \rho \wedge
  (h_- V_3 - h_3 V_-)  \\
  & - \frac{(\theta \cdot V_+)}{2(1- \alpha)(1+
    \alpha)} \rho \wedge (f_- V_3 - f_3 V_-)\nonumber \;.
\end{align}
$d \rho$ can be computed from various combinations of spinor bilinears
(e.g. $d \rho = \tfrac14 (V_+ \lrcorner \Phi_+)$) and we find
\begin{align}
  d \rho & =  \frac{2 \alpha + 1}{1 + \alpha} (\theta \cdot V_+) \rho
  \wedge f + \frac{(\theta \cdot V_+)}{2(1- \alpha)} (f \lrcorner
  \varphi) \wedge V_- \wedge V_3 + \frac{3- 7 \alpha}{1 - \alpha}
  \rho \wedge \left( \tfrac14 h_- V_- + h_3 V_3 \right) \nonumber\\
  &   + \frac{3 (\theta \cdot V_+)}{(1-\alpha)(1+ \alpha)} \rho
  \wedge \left( \tfrac14 f_- V_- + f_3 V_3  \right) + \frac{1+ \alpha}{1- \alpha}
  \varphi \wedge (h_- V_3 - h_3 V_-) \\
  &  + \frac{(\theta \cdot V_+)}{4(1- \alpha)} \varphi \wedge (f_-
  V_3 - f_3 V_-)\nonumber \;.
\end{align}
Again, taking into account the factors which multiply $\rho$ in the
definition of $\Omega$ we find for the latter
\begin{align}
  d \Omega & = \frac32 (\theta \cdot V_+) \Omega \wedge f +
  i \sqrt{\frac2{1 + \alpha}}
  \frac{(\theta \cdot V_+)}{2(1- \alpha)} (f \lrcorner \Omega) \wedge
  V_- \wedge V_3 + \frac{1- 5 \alpha}{1 - \alpha} \Omega \wedge \left( \tfrac14 h_- V_- + h_3
  V_3  \right) \nonumber \\
  & + \frac{3 (\theta \cdot V_+)}{(1-\alpha)(1+ \alpha)} \Omega
  \wedge \left(\tfrac14 f_- V_- + f_3 V_3  \right) + i\sqrt{\frac2{1 + \alpha}}
  \frac{1+ \alpha}{1- \alpha} \Omega \wedge (h_- V_3 - h_3 V_-) \\
  &+ i \sqrt{\frac2{1 + \alpha}} \frac{(\theta \cdot V_+)}{2(1-
    \alpha)(1+ \alpha)} \Omega \wedge (f_- V_3 - f_3 V_-)\nonumber \;.
\end{align}
As in the case of $f=0$ it will be instructive to find the torsion
classes of the manifold with $SU(3)$-structure which is orthogonal to
the vectors. Projecting on the 6-dimensional space the above derivatives become
\begin{align}
  P_6(dJ) & =  - \frac\alpha{1+ \alpha} (\theta \cdot V_+) J \wedge f \;,\\
  P_6(d \Omega) & = \frac32 (\theta \cdot V_+) \Omega \wedge f \;.
\end{align}
We see that now these derivatives are non-zero and in order to read
off the torsion classes of the 6-dimensional manifold with $SU(3)$-structure
we need to first normalise the forms $J$ and $\Omega$. First we normalise $J$ to be a proper almost complex
structure (i.e. its square to be $-1$). This is achieved with the following rescaling
\begin{equation}
  J' = \sqrt{\frac2{1+ \alpha}} J \;,
\end{equation}
Furthermore we define
\begin{equation}
  \varphi' =  \sqrt{\frac2{1 - \alpha}} \varphi \;,
\end{equation}
thus the rescaling of $\rho$ is the following
\begin{equation}
  (\rho')_{mnp} = J'_{rm} (\varphi')_{npr} = \frac2{\sqrt{(1-\alpha)(1+
    \alpha)}} \rho \;.
\end{equation}
We then find the following expressions for the projections of $dJ'$ and $d\varphi'$ on the 6 dimensional space orthogonal to the vectors
\begin{align}
  P_6(d J') & =  - \frac{(\theta \cdot V_+)}2 J' \wedge f \;,\\
  P_6(d \varphi') & =  (\theta \cdot V_+) \varphi' \wedge f \;.
\end{align}
According to \cite{Chiossi:2002tw} this means that the only
non-vanishing torsion classes are $W_4$ and $W_5$ which are given by
\begin{align}
  W_4 & =  \frac12 J' \lrcorner (d J') = - \frac12 (\theta \cdot V_+) f \; , \\
  W_5 & =  \frac12 \varphi' \lrcorner (d \varphi') = (\theta \cdot
  V_+) f \;,
\end{align}
and therefore satisfy $2W_4 + W_5 =0$. 
Note that this relation also featured in the conditions found in
\cite{Lopes Cardoso:2002hd} for non-K\"ahler backgrounds in heterotic
string compactifications.

\section{Summary}
\label{sec:summary}

In this paper we studied $N=2$ compactifications of M-theory to three dimensions which have the defining property that a potential is induced for probe space-time filling M2-branes. Such backgrounds are specifically relevant for many model building applications, ranging from moduli stabilisation to flavour physics, all of which rely on backgrounds that involve potentials for space-time filling probe D3-branes on the F-theory side, which by the F-theory/M-theory duality implies an M2-potential. We showed that it is possible to translate the requirement of an M2-potential to a specific property of the background geometry. Specifically, that the 8-dimensional background should uplift over a trivial circle fibration to a 9-dimensional manifold in which the M2-potential can be related to variations of the angle $\alpha$ between two vectors. In terms of 8-dimensional geometry the variation of $\alpha$ implies that one can define a varying `local' structure group, so that over a generic point the manifold exhibits an 8-dimensional $SU(3)$-structure, while over special loci this changes to $SU(4)$-structure or $G_2$-structure.

We studied the supersymmetry equations over such backgrounds and wrote them as differential constraints on the 9-dimensional forms. We identified a specific 4-form flux that sources variations of the angle $\alpha$ over the space, and showed that it vanishes over the special $SU(4)$-structure and $G_2$-structure loci, while over the generic $SU(3)$-structure locus it is parameterised by two real one-forms $h$ and $g$. We went on to study backgrounds which support this particular 4-form flux as well as a further possible four-form flux with one leg in the internal directions, but no other 4-form fluxes. We showed that in this restricted case, over the generic $SU(3)$-structure locus the background takes the form of a 6-dimensional $SU(3)$-structure manifold with torsion classes $W_4$ and $W_5$, satisfying $2W_4=-W_5$, fibered over a 2-dimensional base. In the case where the additional 1-leg flux is turned off we showed that the geometry is a 6-dimensional Calabi-Yau fibered over a 2-dimensional base which supports the flux and over which $\alpha$ varies. Since a major motivation for our work is an application to F-theory, and this requires that the background supports an elliptic fibration, it is encouraging that the simplest solutions are based on a Calabi-Yau fibration since it is well known how to construct elliptically fibered Calabi-Yau threefolds.

The analysis of the supersymmetry equations performed in this work can form a guide for finding full explicit solutions. This would involve also solving the equations of motion for the supergravity fields, the Bianchi identities, and possibly, if they are not automatically implied, Einstein's equations. It is likely that imposing the full set of requirements for a stable solution, and possibly a realistic vacuum, would imply the need to also incorporate further fluxes, perhaps the analogs of the primitive $(2,2)$ flux. Note that the backgrounds that arise in the simplified flux cases discussed above are naturally similar to the backgrounds studied in \cite{Grimm:2013fua}, and the approach presented in that work of a two-stage reduction may be useful for finding full solutions.

As discussed in the introduction, M2 potentials can be sourced by non-perturbative effects and it would be interesting, perhaps along the lines of \cite{Baumann:2009qx,Baumann:2010sx}, to develop a map between such non-perturbative effects and the flux presented in this work. A related extension of our work would be a more detailed understanding of the full form of the potential that is induced for the M2 branes. Along the same lines, a more detailed study of the applications to the phenomenological aims presented: moduli stabilisation, inflation, flavour physics, would be interesting.

\subsubsection*{Acknowledgments}

We would like to thank James Sparks, Calin Lazaroiu, Mirela Babalic
and Ioana Coman for useful discussions. The work of EP is supported by
the Heidelberg Graduate School for Fundamental Physics. The work of AM
was partly supported by the UEFISCDI grant TE\_93, 77/4.08.2010 and by
the ``Nucleu'' project.

\appendix

\section{Fierz identities}
\label{sec:frz}

Fierz relations represent identities of products of gamma matrices
with reshuffled indices. They emerge as a direct consequence of the
the fact that the elements of the Clifford algebra in $d$-dimensions
form a basis for the square matrices $(2^{\left[\tfrac{d}2 \right]}
\times 2^{\left[\tfrac{d}2\right]})$. Thus any square matrix $M$ can be expanded in a basis of
gamma matrices in the following way
\begin{align}
M = \frac{1}{2^{\left[\tfrac{d}2 \right]}} \sum_A \textrm{Tr}(M \gamma^A) \gamma_A
\label{gammabasis}
\end{align}
Fierz identities take a simpler form in the case of Majorana spinors. We restrict to this case and assume that
the gamma matrices are real and symmetric. In order to write the general quadratic Fierz identity one can define
the following matrix
\begin{equation}
M_{cd} = (\gamma_A)_{ac} (\gamma_B)_{bd}
\end{equation}
for arbitrary fixed spinorial indices $a,b$ and use eq. \eqref{gammabasis} to obtain
\begin{equation}
(\gamma_A)_{ac} (\gamma_B)_{bd} = \frac{1}{2^{\left[\tfrac{d}2 \right]}} \sum_C (\gamma_A \gamma_C \gamma_B^T)_{ab}(\gamma^C)_{cd}
\label{generalF}
\end{equation}
The equation above can be used to generate all the necessary Fierz identities. For instance, by taking $\gamma_A =\gamma_B = \mathbbm{1}$ one obtains the well-known completeness relation for gamma matrices. Above, we have chosen to reshuffle the indices $b$ and $c$. Similarly one can obtain relations with other indices reshuffled.

We shall use the tensor form of
the Fierz identities in eq. \eqref{generalF} which is obtained by contracting with the invariant spinors
on the nine-dimensional manifold. The relations we obtain are exhaustive as $\gamma_A, \gamma_B$ run over all the basis elements of the  Clifford algebra.

An equivalent approach is to start with the completeness relation for the gamma matrices
\begin{equation}
(\mathbbm{1})_{ac} (\mathbbm{1})_{bd} = \frac{1}{2^{\left[\tfrac{d}2 \right]}} \sum_C ( \gamma_C )_{ab}(\gamma^C)_{cd}
\label{completeness}
\end{equation}
and contract with arbitrary spinors. In our case the spinors are chosen as the invariant spinors
on the nine-dimensional manifold multiplied by arbitrary elements of the
Clifford algebra. These spinors do not form a basis on the space of
spinors as they are not all linearly independent, but it is clear that they form a generating set.  Therefore, the relations we
obtain are exhaustive and they are equivalent to the original statement that the gamma matrices form a basis.

In this appendix we summarize the results obtained by performing a
linear analysis of the system of equations generated in the way
described above.  We list the Fierz identities according to the
number of space-time free indices. We also split the results into
identities which have completely antisymmetric free indices and the
ones which have symmetries corresponding to other Young tableaux. For
the antisymmetric identities, the maximum number of free indices which
produces new results is four, as other relations with more
antisymmetric free indices can be obtained by contracting with the
nine-dimensional $\epsilon$ symbol. For identities which have other
symmetries of the free indices, we also stop at four indices as in our
calculations we do not need further relations.

\subsection{Completely antisymmetric Fierz identities}
\subsubsection*{No free indices}
\begin{align}
&||V_1||^2 = 1   && \langle V_1, V_2\rangle = \alpha\\
&||V_2||^2 = 1  && \langle V_1, V_3\rangle = 0\\
&||V_3||^2 = \frac{1}{2}(\, 1- \alpha \, )  && \langle V_2, V_3\rangle = 0
\end{align}
\vspace{-0.7cm}
\begin{align}
&||K||^2 = \frac{1}{2}(5+ 3\alpha )\\
&||\Psi||^2 = \frac{1}{2}(11- 3\alpha)
\end{align}
\vspace{-0.7cm}
\begin{align}
& ||\Phi_1||^2 = 14 && \langle \Phi_1, \Phi_2 \rangle = -1 - \alpha\\
&||\Phi_2||^2 = 14  && \langle \Phi_1, \Phi_3 \rangle =0 \\
&||\Phi_3||^2 = \frac{1}{2}(15+ \alpha ) && \langle \Phi_2, \Phi_3 \rangle =0
\end{align}
Notice that the real parameter $\alpha$ can take values only in the interval
\begin{align}
\alpha \in [-1,1]
\end{align}
%

\subsubsection*{One free index}
%
%
\begin{align}
&V_1^m K_m{}^r=V_3^r  && \frac{1}{3!} \Psi^{mnp} (\Phi_1)_{mnp}{}^r=7 V_3^r\\
&V_2^m K_m{}^r=-V_3^r  && \frac{1}{3!} \Psi^{mnp} (\Phi_2)_{mnp}{}^r=-7 V_3^r\\
&V_3^m K_m{}^r=-\frac{1}{2}(V_1^r-V_2^r) && \frac{1}{3!} \Psi^{mnp} (\Phi_3)_{mnp}{}^r=- \frac{7}{2} (V_1^r-V_2^r)
\end{align}
\begin{equation}
\frac{1}{2!} K^{mn} \Psi_{mn}{}^r=2(V_1^r+V_2^r)
\end{equation}
\begin{align}
&\frac{1}{4!}(\Phi_1)^{mnpq} (*\Phi_1)_{mnpq}{}^r = 14 V_1^r  && \frac{1}{4!}(\Phi_1)^{mnpq} (*\Phi_2)_{mnpq}{}^r = -(V_1^r+V_2^r)\\
&\frac{1}{4!}(\Phi_2)^{mnpq} (*\Phi_2)_{mnpq}{}^r = 14 V_2^r  && \frac{1}{4!}(\Phi_1)^{mnpq} (*\Phi_3)_{mnpq}{}^r = 7 V_3^r\\
&\frac{1}{4!}(\Phi_3)^{mnpq} (*\Phi_3)_{mnpq}{}^r = 4(V_1^r+V_2^r) && \frac{1}{4!}(\Phi_2)^{mnpq} (*\Phi_3)_{mnpq}{}^r = 7 V_3^r
\end{align}
%
%

\subsubsection*{Two free indices}
\begin{align}
&V_1^m\Psi_m{}^{rs}= K^{rs}-2V_1^{[r}V_3^{s]} && \frac{1}{2} K^{mn} (\Phi_1)_{mn}{}^{rs} =-3K^{rs}+6 V_1^{[r}V_3^{s]}\\
&V_2^m\Psi_m{}^{rs}= K^{rs}+2V_2^{[r}V_3^{s]} && \frac{1}{2}K^{mn} (\Phi_2)_{mn}{}^{rs} =-3K^{rs}-6V_2^{[r}V_3^{s]}\\
&V_3^m\Psi_m{}^{rs}= -V_1^{[r}V_2^{s]} && \frac{1}{2}K^{mn} (\Phi_3)_{mn}{}^{rs} =3V_1^{[r}V_2^{s]}
\end{align}
\begin{align}
\frac{1}{3!}\Psi^{mnp}(* \Phi_1)_{mnp}{}^{rs} &=-3K^{rs}-8 V_1^{[r}V_3^{s]}\\
\frac{1}{3!}\Psi^{mnp}(* \Phi_2)_{mnp}{}^{rs}&=-3K^{rs}+8 V_2^{[r}V_3^{s]}\\
\frac{1}{3!}\Psi^{mnp}(* \Phi_3)_{mnp}{}^{rs}&=-4V_1^{[r}V_2^{s]}
\end{align}
\vspace{-0.7 cm}
\begin{align}
&\frac{1}{3!}(\Phi_1)^{mnp[r}(\Phi_1)_{mnp}{}^{s]} = 0   && \frac{1}{3!}(\Phi_1)^{mnp[r}(\Phi_2)_{mnp}{}^{s]} =-V_1^{[r}V_2^{s]}\\
&\frac{1}{3!}(\Phi_2)^{mnp[r}(\Phi_2)_{mnp}{}^{s]} = 0   && \frac{1}{3!}(\Phi_1)^{mnp[r}(\Phi_3)_{mnp}{}^{s]} =4 K^{rs}- V_1^{[r}V_3^{s]}\\
&\frac{1}{3!}(\Phi_3)^{mnp[r}(\Phi_3)_{mnp}{}^{s]} = 0   && \frac{1}{3!}(\Phi_2)^{mnp[r}(\Phi_3)_{mnp}{}^{s]} = -4 K^{rs}- V_2^{[r}V_3^{s]}
\end{align}
\begin{align}
&K^{m[r}K_m{}^{s]}=0\\
&\Psi^{mn[r}\Psi_{mn}{}^{s]}=0
\end{align}


\subsubsection*{Three free indices}
\begin{align}
&V_1^m(\Phi_1)_m{}^{rst}=0  && V_2^m (\Phi_1)_m{}^{rst}=6V_3^{[r}K^{st]}+2K^{m[r}\Psi_m{}^{st]}\\
&V_1^m (\Phi_2)_m{}^{rst}=-6V_3^{[r}K^{st]}  +2 K^{m[r}\Psi_m{}^{st]} &&  V_2^m(\Phi_2)_m{}^{rst}=0 \\
&V_1^m (\Phi_3)_m{}^{rst}=\Psi^{rst}-3V_1^{[r}K^{st]} && V_2^m (\Phi_3)_m{}^{rst}=-\Psi^{rst}+3V_2^{[r}K^{st]}
\end{align}
\begin{align}
&V_3^m(\Phi_1)_m{}^{rst}=-\Psi^{rst}+3V_1^{[r}K^{st]}  && \frac{1}{2}K^{mn}(*\Phi_1)_{mn}{}^{rst} = -\Psi^{rst}-6V_1^{[r}K^{st]}\\
&V_3^m(\Phi_2)_m{}^{rst}=\Psi^{rst}-3V_2^{[r}K^{st]}   &&  \frac{1}{2}K^{mn}(*\Phi_2)_{mn}{}^{rst} = -\Psi^{rst}-6V_2^{[r}K^{st]}\\
&V_3^m (\Phi_3)_m{}^{rst}=-K^{m[r}\Psi_m{}^{st]}  &&  \frac{1}{2}K^{mn}(*\Phi_3)_{mn}{}^{rst}=-6V_3^{[r}K^{st]}
\end{align}
\begin{align}
&\frac{1}{2}\Psi^{mn[r}(\Phi_1)_{mn}{}^{st]} = -2\Psi^{rst}+3V_1^{[r}K^{st]}\\
&\frac{1}{2}\Psi^{mn[r}(\Phi_2)_{mn}{}^{st]} = -2\Psi^{rst}+3V_2^{[r}K^{st]}\\
&\frac{1}{2}\Psi^{mn[r}(\Phi_3)_{mn}{}^{st]} = 3 V_3^{[r}K^{st]}
\end{align}
\begin{align}
&\frac{1}{3!}(\Phi_1)^{mnp[r}(*\Phi_1)_{mnp}{}^{st]}=0   && \frac{1}{3!} (\Phi_1)^{mnp[r} (*\Phi_2)_{mnp}{}^{st]}=4V_3^{[r}K^{st]} \\
&\frac{1}{3!}(\Phi_2)^{mnp[r}(*\Phi_2)_{mnp}{}^{st]}=0   && \frac{1}{3!} (\Phi_1)^{mnp[r} (*\Phi_3)_{mnp}{}^{st]}= 2\Psi^{rst}+2V_1^{[r}K^{st]} \\
&\frac{1}{3!}(\Phi_3)^{mnp[r}(*\Phi_3)_{mnp}{}^{st]}=0   && \frac{1}{3!} (\Phi_2)^{mnp[r} (*\Phi_3)_{mnp}{}^{st]}= -2\Psi^{rst}-2V_1^{[r}K^{st]}
\end{align}
\begin{align}
&\frac{1}{3!} (\Phi_2)^{mnp[r} (*\Phi_1)_{mnp}{}^{st]}=-4V_3^{[r}K^{st]}\\
&\frac{1}{3!}(\Phi_3)^{mnp[r} (*\Phi_1)_{mnp}{}^{st]}=-2\Psi^{rst}-2V_1^{[r}K^{st]}\\
&\frac{1}{3!}(\Phi_3)^{mnp[r} (*\Phi_2)_{mnp}{}^{st]}=2\Psi^{rst}+2V_2^{[r}K^{st]}
\end{align}


\subsubsection*{Four free indices}
\begin{align}
&V_1^m(*\Phi_1)_m{}^{rsuv}=(\Phi_1)^{rsuv}\\
&V_1^m(*\Phi_2)_m{}^{rsuv} =-(\Phi_1)^{rsuv}-8V_3^{[r}\Psi^{suv]}-6K^{[rs}K^{uv]}\\
&V_1^m (*\Phi_3)_m{}^{rsuv} = (\Phi_3)^{rsuv}-4 V_1^{[r}\Psi^{suv]}
\end{align}
\begin{align}
&V_2^m(*\Phi_1)_m{}^{rsuv}=-(\Phi_2)^{rsuv}+8V_3^{[r}\Psi^{suv]}-6K^{[rs}K^{uv]}\\
&V_2^m(*\Phi_2)_m{}^{rsuv}=(\Phi_2)^{rsuv}\\
&V_2^m (*\Phi_3)_m{}^{rsuv} = (\Phi_3)^{rsuv}+4 V_2^{[r}\Psi^{suv]}
\end{align}
\begin{align}
&V_3^m (*\Phi_1)_m{}^{rsuv} = 4V_1^{[r}\Psi^{suv]}   \\
&V_3^m (*\Phi_2)_m{}^{rsuv} =- 4V_2^{[r}\Psi^{suv]}  \\
&V_3^m (*\Phi_3)_m{}^{rsuv}=\frac{1}{2}(\Phi_1+\Phi_2)^{rsuv}+3K^{[rs}K^{uv]}
\end{align}
\begin{align}
& K^{m[r}(\Phi_1)_m{}^{suv]}= (\Phi_3)^{rsuv}-3 V_1^{[r}\Psi^{suv]} \\
& K^{m[r}(\Phi_2)_m{}^{suv]}= -(\Phi_3)^{rsuv}-3 V_2^{[r}\Psi^{suv]}\\
& K^{m[r}(\Phi_3)_m{}^{suv]}=-\frac{1}{2}(\Phi_1-\Phi_2)^{rsuv}-3V_3^{[r}\Psi^{suv]}
\end{align}
\begin{align}
&\frac{1}{2} \Psi^{mn[r}(*\Phi_1)_{mn}{}^{suv]}=(\Phi_3)^{rsuv}+2 V_1^{[r}\Psi^{suv]}\\
&\frac{1}{2} \Psi^{mn[r}(*\Phi_2)_{mn}{}^{suv]}=-(\Phi_3)^{rsuv}+2 V_2^{[r}\Psi^{suv]}\\
&\frac{1}{2}\Psi^{mn[r}(*\Phi_3)_{mn}{}^{suv]}=-\frac{1}{2}(\Phi_1-\Phi_2)^{rsuv}+2V_3^{[r}\Psi^{suv]}
\end{align}
\begin{align}
&\frac{1}{2}K^{mn}(*\Psi)_{mn}{}^{rsuv}=-\frac{1}{2}(\Phi_1+\Phi_2)^{rsuv}+3K^{[rs}K^{uv]} \\
&\Psi^{m[rs}\Psi_m{}^{uv]}=-\frac{1}{2}(\Phi_1+\Phi_2)^{rsuv}-2K^{[rs}K^{uv]}
\end{align}
\begin{align}
&\frac{1}{2}(\Phi_1)^{mn[rs}(\Phi_1)_{mn}{}^{uv]}=-2(\Phi_1)^{rsuv}  && \frac{1}{2}(\Phi_1)^{mn[rs}(\Phi_2)_{mn}{}^{uv]}=-2K^{[rs}K^{uv]}\\
&\frac{1}{2}(\Phi_2)^{mn[rs}(\Phi_2)_{mn}{}^{uv]}=-2(\Phi_2)^{rsuv}   && \frac{1}{2} (\Phi_1)^{mn[rs}(\Phi_3)_{mn}{}^{uv]}=-(\Phi_3)^{rsuv}\\
&\frac{1}{2}(\Phi_3)^{mn[rs}(\Phi_3)_{mn}{}^{uv]}=-\frac{1}{2}(\Phi_1+\Phi_2)^{rsuv}+K^{[rs}K^{uv]}  && \frac{1}{2} (\Phi_2)^{mn[rs}(\Phi_3)_{mn}{}^{uv]}=-(\Phi_3)^{rsuv}
\end{align}


\subsection{Fierz identities with symmetric part}
We list here the Fierz identities which can have a symmetric part, that is the ones which lie in the tensorial algebra.  We restrict only to necessary identities, that is the ones involving the forms $V_3,\ K,\ \Psi\ $ and $\ \Phi_3$. The rest can be obtained from these ones making also use of the antisymmetric identities already given earlier in the appendix.
\subsubsection*{Two indices}

\begin{align}
&K^{mr}K_m{}^s=\frac{1}{2}(1+\alpha)\delta^{rs} - V_1^{(r}V_2^{s)}+V_3^rV_3^s \label{symK}\\
&\Psi^{mnr}\Psi_{mn}{}^s=(4-2\alpha)\delta^{rs}+6V_1^{(r}V_2^{s)}-6V_3^rV_3^s \label{symPsi}\\
&(\Phi_3)^{mnpr}(\Phi_3)_{mnp}{}^s=3(7+\alpha)\delta^{rs}-24V_1^{(r}V_2^{s)}-18V_3^rV_3^s
\end{align}

\subsubsection*{Three indices}

\begin{align}
&K^{mr}\Psi_m{}^{st}=\delta^{r[s}(V_1^{t]}+V_2^{t]})+K^{m[r}\Psi_m{}^{st]}\\
&\Psi^{mnr}(\Phi_3)_{mn}{}^{st}=-6\delta^{r[s}(V_1^{s]}-V_2^{s]})+12V_3^{[r}K^{st]}-6V_3^r K^{st}\\
&\frac{1}{3!}(\Phi_3)^{mnpr}(*\Phi_3)_{mnp}{}^{st}=4\delta^{r[s}(V_1^{t]}+V_2^{t]})
\end{align}

\subsubsection*{Four indices}

\begin{align}
\Psi^{mrs}\Psi_{muv}&=(1-\alpha)\delta^{rs}_{uv}+2\delta^{[r}_{[u}\left(V_1^{s]}V_{2v]}+V_{1v]}V_2^{s]}-2V_3^{s]}V_{3v]}\right)\\
 &-\frac{1}{2}(\Phi_1+\Phi_2)^{rs}{}_{uv}-3K^{[rs}K^{uv]}+K^{rs}K_{uv}\\
\frac{1}{2}(\Phi_3)^{mnrs}(\Phi_3)_{mnuv}&=(3+\alpha)\delta^{rs}_{uv}-4\delta^{[r}_{[u}\left(V_1^{s]}V_{2v]}+V_{1v]}V_2^{s]}+V_3^{s]}V_{3v]}\right)\\
&-\frac{1}{2}(\Phi_1+\Phi_2)^{rs}{}_{uv}+3K^{[rs}K^{uv]}-2K^{rs}K_{uv}\\
\end{align}

\section{Relations satisfied by the $SU(3)$ structure forms}
\label{sec:SU3app}
In the main text we derived the parametrisation of the spinor
bilinears \eqref{bilinears} in terms of forms defining a SU(3)
structure in six dimensions and three additional vectors $V_\pm$ and
$V_3$. Here we shall give more details about the relations which these
forms satisfy, including also brief indications on how to derive such
relations. The crucial relation which we shall use almost everywhere
is the symmetric relation obtained by contracting $J$ with itself over
one index which can be derived from \eqref{symK}
\begin{align}
  \label{J2}
  J_{mn} J^n{}_p & =  -\frac12 (1+ \alpha) \delta_{mp} + \frac14 (V_+)_m
  (V_+)_p + \frac{1+\alpha}{4(1-\alpha)}(V_-)_m (V_-)_p +
  \frac{1+\alpha}{1-\alpha}(V_3)_m (V_3)_p \nonumber \\
   & =  \frac12 (1+ \alpha) \left( - \delta_{mp} + (P_+)_{mp} +
     (P_-)_{mp} + (P_3)_{mp} \right) \; .
\end{align}
In the main text we claimed that $\rho$ defined as in \eqref{rhodef}
is totally antisymmetric. Using the Fierz relations (with 3 free
indices) which give the contractions of the vectors $V_{1,2}$ with the
forms $\Phi_{1,2}$ we can construct the object $V_+ \lrcorner
\Phi_+$. Using \eqref{Phi+par} we find
\begin{equation}
  J_{mn} \varphi^m{}_{pq} = K_{m[n} \Psi^m{}_{pq]} \; ,
  \label{antirho}
\end{equation}

Let us continue by computing the norms of the SU(3) forms. Note that
since we have done an orthogonal decomposition in terms of the vector
fields, the terms on the RHS of \eqref{Kpar}--\eqref{Phi3par} are
independent, in the sense that total contractions of different terms
vanish by definition. Using the norms of the vector fields which were
listed at the beginning of this appendix, we can immediately derive
the norm of $J$ as
\begin{equation}
  \label{nJ}
  J \lrcorner J = \tfrac12 J_{mn} J^{mn} = \tfrac32 (1+ \alpha)
\end{equation}
Taking the square of \eqref{Psipar} we find in  a similar way the norm
of $\varphi$
\begin{equation}
  \label{nphi}
  \varphi \lrcorner \varphi = \tfrac16 \varphi_{mnp} \varphi^{mnp} = 2
  (1- \alpha) \; .
\end{equation}
From \eqref{rhodef} and using \eqref{J2} we find
\begin{equation}
  \label{nr}
  \rho \lrcorner \rho = \tfrac16 \rho_{mnp} \rho^{mnp} = (1-\alpha)
  (1+ \alpha)
\end{equation}
Using the Fierz relation involving the contraction of $K$ and $\Psi$
over two indices, we immediately find
\begin{equation}
  J \lrcorner \varphi = 0 = J \lrcorner \rho \;
\end{equation}
where the second equality holds due to the fact that $\rho$ in
\eqref{rhodef} is totally antisymmetric. Using the above relation and
the Fierz identity involving the contraction of $\Psi$ and
$\Phi_{1,2}$ over three indices, we obtain
\begin{equation}
  \varphi \lrcorner \rho = 0 \; .
\end{equation}

Most of the other relations we shall need involve a Hodge $*$
operation and are somehow more complicated. Let us look at the Fierz
relation which gives the contraction of $\Phi_3$ with $* \Phi_3$ over
four indices. This can be rewritten in form notation as
\begin{equation}
  *(\Phi_3 \wedge \Phi_3) = 4 V_+\; .
\end{equation}
Using \eqref{Phi3par} we find
\begin{equation}
  \frac4{(1- \alpha)^2} \varphi \wedge \rho \wedge V_- \wedge V_3 -
  \frac2{(1-\alpha)(1+\alpha)} \rho \wedge J \wedge V_+ \wedge V_-
  \wedge V_3 = 4 * V_+ \; .
\end{equation}
Clearly, the second term on the LHS must vanish, as it contains $V_+$
while on the RHS we only find $*V_+$. $J$ and $\rho$ are orthogonal to
the vectors, and the only way this term can vanish is if
\begin{equation}
  \rho \wedge J = 0 \; .
\end{equation}
The remaining relation can be rewritten, by contracting $V_-$ and
$V_3$, as
\begin{equation}
  \varphi\wedge \rho = *(V_+ \wedge V_- \wedge V_3) \; .
\end{equation}
From Fierz identities involving the contraction of $\Phi_{1,2}$ with
$\Phi_3$ we find similar relations, and again, by contracting the
appropriate vectors we find
\begin{equation}
  J \wedge J \wedge J = \frac32 \; \frac{1+ \alpha}{1- \alpha} *(V_+
  \wedge V_- \wedge V_3) \; .
\end{equation}
Further relations with Hodge star can be derived form these ones by
contracting with appropriate forms and making use of the orthogonality
conditions and of the norms of the various quantities. We find
\begin{align}
 *J &= \frac{1}{2(1+\alpha)(1-\alpha)}J\wedge J \wedge V_+ \wedge V_- \wedge V_3\\
 *\varphi &= \frac{1}{(1+\alpha)(1-\alpha)} \rho \wedge V_+ \wedge V_- \wedge V_3\\
 *\rho &= -\frac{1}{2(1+\alpha)} \varphi \wedge V_+ \wedge V_- \wedge V_3
\end{align}
Other useful relations which can be derived easily from the ones already written so far are
\begin{align}
 *(J \wedge V_+ \wedge V_- \wedge V_3) & =  (1-\alpha) J \wedge J \\
 *(J \wedge V_- \wedge V_3) & =  \frac{(1-\alpha)}{2(1+ \alpha)} J
 \wedge J \wedge V_+\\
 *(J \wedge V_+ \wedge V_3) & =  - \frac12J \wedge J \wedge V_-\\
 *(J \wedge V_+ \wedge V_-) & =  2J \wedge J
 \wedge V_3
\end{align}
and
\begin{align}
 *(\rho \wedge V_+ \wedge V_3) & =  \frac12 (1+ \alpha)\varphi \wedge
 V_- &  *(\rho \wedge V_3) & =  \frac14 \varphi \wedge V_+ \wedge V_- \\
 *(\rho \wedge V_- \wedge V_3) & =  - \frac12 (1- \alpha)\varphi \wedge
 V_+ & *(\rho \wedge V_+) & =  \frac{1+ \alpha}{1- \alpha} \varphi \wedge
  V_- \wedge V_3 \\
 *(\rho \wedge V_+ \wedge V_-) & =  - 2(1+ \alpha) \varphi \wedge V_3
 & *(\rho \wedge V_-) & =  - \varphi \wedge V_+ \wedge V_3
\end{align}
One can also compute the Hodge duals of the original spinor bilinear forms and express them in terms of the $SU(3)$ parametrisation
\begin{align}
  *K &= \frac{1}{2(1+\alpha)(1-\alpha)} J\wedge J \wedge V_+\wedge V_- \wedge V_3+ \frac{1}{2(1+\alpha)(1-\alpha)} \varphi\wedge\rho\wedge V_+\\
 *\Psi & = \frac{1}{(1+\alpha)(1-\alpha)} \rho\wedge V_+\wedge V_- \wedge V_3+\frac{1}{(1+\alpha)(1-\alpha)} J\wedge J\wedge V_-\wedge V_3+\frac{1}{2(1-\alpha)}\varphi\wedge\rho\\
 *\Phi_+ &= -\frac{2}{(1+\alpha)(1-\alpha)} J \wedge V_+ \wedge V_- \wedge V_3 -\frac{2}{1-\alpha} \varphi \wedge V_- \wedge V_3-\frac{1}{1+\alpha} J\wedge J \wedge V_+\\
 *\Phi_- &= -\frac{2}{(1+\alpha)(1-\alpha)}\rho\wedge V_+ \wedge V_- -\frac{2}{1-\alpha}\varphi\wedge V_+\wedge V_3-\frac{1}{1+\alpha} J\wedge J \wedge V_- \\
 *\Phi_3 & = -\frac{2}{(1+\alpha)(1-\alpha)} \rho\wedge V_+ \wedge V_3+\frac{1}{2(1-\alpha)} \varphi\wedge V_+ \wedge V_- -\frac{1}{1+\alpha} J\wedge J \wedge V_3
\end{align}

From the symmetric Fierz identity \eqref{symPsi} involving the contraction of $\Psi$
with itself over one index we can derive a similar relation for
$\varphi$
\begin{align}
\label{phiphi}
  \varphi^{mrs} \varphi_{mtu} & =  (1-\alpha) \delta^{rs}_{tu} + 2
  \frac{1-\alpha}{1+ \alpha} J^r{}_{[t} J_{u]}{}^s  - \frac{1-\alpha}{1+ \alpha} \delta^{[r}_{[t} V_{+u]} V_+^{s]} -
  \delta^{[r}_{[t} V_{-u]} V_-^{s]} -4\delta^{[r}_{[t} V_{3u]}
  V_3^{s]} \nonumber \\
 & + \frac1{2(1+\alpha)} V_{+[t} V_{-u]} V_+ ^{[r} V_-^{s]}
 + \frac2{1+\alpha} V_{+[t} V_{3u]} V_+ ^{[r} V_3^{s]}
   + \frac2{1-\alpha} V_{-[t} V_{3u]} V_- ^{[r} V_3^{s]}  \; .
\end{align}
By contracting this relation with $J$ we can find similar relations
for $\rho$ or $\rho$ and $\varphi$
\begin{equation}
  \rho_m{}^{rs} \varphi^m{}_{tu}= 2 (1-\alpha) \delta_{[t}^{[r}
  J_{u]}{}^{s]} + \frac{1- \alpha}{1+ \alpha} J^{[r}{}_{[t}
  V_{+u]} V_+^{s]} + J^{[r}{}_{[t} V_{-u]} V_-^{s]} + 4 J^{[r}{}_{[t}
  V_{3u]} V_3^{s]}
\end{equation}
\begin{align}
\label{rhorho}
  \rho^{mrs} \rho_{mtu} & =  \tfrac12 (1-\alpha)(1+ \alpha)
  \delta^{rs}_{tu}
  + (1-\alpha) J^r{}_{[t} J_{u]}{}^s  - \frac12(1-\alpha)
  \delta^{[r}_{[t} V_{+u]} V_+^{s]} - \frac12 (1+ \alpha)
  \delta^{[r}_{[t} V_{-u]} V_-^{s]} \nonumber \\
  & - 2 (1+ \alpha) \delta^{[r}_{[t} V_{3u]}
  V_3^{s]}  + \tfrac14 V_{+[t} V_{-u]} V_+ ^{[r} V_-^{s]}
  + V_{+[t} V_{3u]} V_+ ^{[r} V_3^{s]} + \frac{1+ \alpha}{1-\alpha}
  V_{-[t} V_{3u]} V_- ^{[r} V_3^{s]} \; .
\end{align}
Finally, by contracting a pair of indices in the above relations it is easy to find
\begin{eqnarray}
  \varphi_{mnr} \varphi^{mns} & = & 2 (1- \alpha) \delta^s_r -
  \frac{1-\alpha}{1+\alpha} V_{+r}  V_+^s   - V_{-r}
  V_-^s - 4 V_{3r} V_3^s \; ; \nonumber \\
  \varphi_{mnr} \rho^{mns} & = & 2 (1- \alpha) J_r{}^{s} \\
  \rho_{mnr} \rho^{mns} & = & (1- \alpha)(1+ \alpha) \delta^s_r -
  \frac{1-\alpha}2 V_{+r} V_+^s - \frac{1+ \alpha}2 V_{-r}
  V_-^s - 2(1+ \alpha) V_{3r} V_3^s \; . \nonumber
\end{eqnarray}

Before we end this section, let us note the following fact which is
useful during the
calculations. Consider a 1-form $k$ which is
orthogonal to all the vector fields, i.e. $k \cdot V_i = 0$, and
define $\tilde k = k \lrcorner J$. By contracting $J$ we find the
equivalent relation
\begin{equation}
  \tilde k \lrcorner J = - \frac{1+ \alpha}2 k \;.
\end{equation}
Furthermore, by contracting with $\varphi$ and taking into account the definition
of $\rho$ from \eqref{rhodef} we find
\begin{equation}
  \label{kcon}
  \tilde k \lrcorner \varphi + k \lrcorner \rho = 0 \; , \qquad
  \frac{1+ \alpha}2 k \lrcorner \varphi - \tilde k \lrcorner \rho = 0
\end{equation}
Taking now the exterior product with $\rho$ and $\varphi$ we find
\begin{eqnarray}
  \tilde k \wedge \varphi + (k \lrcorner \varphi) \wedge J = 0\; ,  & &
  \tilde k \wedge \rho + (k \lrcorner \rho) \wedge J = 0 \nonumber \\
  - (\tilde k \lrcorner \rho) \wedge J + \frac{1+ \alpha}2   k \wedge \rho
  =0\; ,  & & - (\tilde k \lrcorner \varphi) \wedge J + \frac{1+
    \alpha}2 k \wedge \varphi = 0 \nonumber
\end{eqnarray}
where we used the identity $(k \lrcorner \varphi) \wedge J - \varphi
\wedge (k \lrcorner J) = k \lrcorner (\varphi \wedge J) = 0$ and
similar ones for $\rho$ and $\tilde k$. Adding up the equations in the
same column in such a way to obtain the combinations in \eqref{kcon}
we find
\begin{equation}
  \label{kxp}
  \tilde k \wedge \varphi + k \wedge \rho = 0 \; , \qquad \tilde k
  \wedge \rho - \frac{1+ \alpha}2 k \wedge \varphi = 0
\end{equation}
These relations can be intuitively understood in a simple way. From
the 1-forms $k$ and $\tilde k$ we can construct a complex $(1,0)$ form
\begin{equation}
  \textgoth{k} = k + i \sqrt{\frac2{1+\alpha}} \tilde k
\end{equation}
Then the expressions in \eqref{kxp} are nothing but the real and
imaginary components of the (4,0) form $\textgoth{k} \wedge
\Omega$. But this form should vanish identically since it only lives on
the 6-dimensional space orthogonal to the vectors and this gives the relations in
\eqref{kxp}.


\section{Supersymmetry equations}
\label{sec:susyeqapp}

We summarize in this appendix the supersymmetry algebraic constraints arising from the variation of the external components of the
gravitino. For the $N=2$ flux background that we consider one has to satisfy the equations
\begin{align}
Q\xi_j=0 \qquad j=1,2 \label{twoQ}
\end{align}
where the operator $Q$ is given in eq. \eqref{Q9d} for the auxiliary 9d manifold $Y_9$. We translate the equations above into constraints
on the fluxes $\tilde f$ and $F$ involving the spinor bilinears defined in eq. \eqref{bilinears}. Specifically, we contract eq. \eqref{twoQ}
with the following generating set of the spinorial representation
\begin{align}
\gamma_A \xi_i \qquad i=1,2 \qquad \textrm{and} \qquad \gamma_A \in \left\{\mathbbm{1},\gamma_m, \gamma_{mn},\gamma_{mnp},\gamma_{mnpq}\right\}
\end{align}
It is then convenient to represent the algebraic constraints in eq. \eqref{twoQ} in the the following equivalent form
\begin{align}
\xi_i^T \left[Q\gamma_A \pm \gamma_A^T Q^T\right]\xi_j = 0 \qquad i,j=1,2 \label{algebraic}
\end{align}
After inserting the expression of $Q$ given in eq. \eqref{Q9d} and expanding the products of gamma matrices one can express the resulting equations in terms of the spinor bilinears in eq. \eqref{bilinears}. The result is the following (a number of these expressions first appeared in \cite{Lazaroiu:2012fw})
\begin{align}
  \label{Qconstraints}
  - \lambda (\theta \cdot V_i) + \frac12 (V_i \cdot d \Delta) -
  \frac1{12} F \lrcorner \Phi_i = 0 \; ; \qquad \tilde f_m \theta_n K^{mn}
  & = 0 \; \\
  - 2 \lambda \theta + d \Delta - \frac1{12} *(F \wedge \Phi_+)
  + \frac16 V_+ \lrcorner (\tilde f \wedge \theta) & =  0 \; \\
  - \frac1{12} *(F \wedge \Phi_{-\;,3}) + \frac16 V_{- \; , 3}
  \lrcorner (\tilde f \wedge \theta) & =  0 \; \\
  2 \lambda \theta \lrcorner K - d \Delta \lrcorner K + \frac16 \Psi
  \lrcorner F - \frac13(\tilde f \wedge \theta) \lrcorner \Psi & =  0 \\
  2 \lambda \theta_{[m} V_{+\; n]} - \partial_{[m} \Delta V_{+\; n]}
  + \frac1{36} F_{[m}{}^{pqr} \Phi_{+\; n] pqr} - \frac16 \Phi_{+\;
    mnpq}\tilde f^p \theta^q + \frac23 \tilde f_{[m} \theta_{n]} & = 0 \qquad \\
  2 \lambda \theta_{[m} (V_{-\; 3})_{n]} - \partial_{[m} \Delta
   (V_{-\; 3})_{n]} + \frac{1}{36}\, F_{[m}{}^{klp} (\Phi_{- \; 3})_{ n]klp}
   - \frac16 (\Phi_{- \; 3})_{mnkl} \tilde f^k \theta^l & =  0 \\
  - 2 \lambda \theta \lrcorner \Psi + d \Delta \lrcorner \Psi
   + \frac16 *(F \wedge \Psi) + \frac16 K \lrcorner F + \frac13 (\theta
   \lrcorner K) \wedge \tilde f - \frac13 (\tilde f \lrcorner K) \wedge \theta & =
   0 \\
  - \lambda \theta_k \Phi_{i \; mnpk} + \frac12 \partial_k\Delta
   \Phi_{i\; mnpk} - \frac1{24}\, F_{[p}{}^{klq} (*\Phi_i)_{mn]klq}
   + \frac1{12} F_{mnpk} V_i^k &  \nonumber \\
  - \frac16 *(\Phi_i \wedge \tilde f \wedge
   \theta)_{mnp} + \frac16 (V_i \wedge \tilde f \wedge \theta)_{mnp} & =  0 \\
   - 2 \lambda (K \wedge \theta)_{mnp} + K \wedge d \Delta_{mnp}
   + \frac16 *(F \wedge K)_{mnp}  + \frac14 \Psi_{[m}{}^{kl}
   F_{np]kl} &  \nonumber \\
   - \frac13 [(\theta \lrcorner \Psi) \wedge \tilde f]_{mnp} +\frac13 [(\tilde f \lrcorner \Psi) \wedge
   \theta]_{mnp}\ & =  0 \\
   - \lambda *(\Phi_+ \wedge \theta)_{mnpq} + \frac12  *(\Phi_+ \wedge
   d \Delta)_{mnpq} - \frac1{12} *(F \wedge V_+)_{mnpq}   & \nonumber \\
   + \frac14 F_{[mn}{}^{rs} (\Phi_+)_{pq]rs}
   - \frac16 F_{mnpq} + \frac16 (\theta
   \lrcorner \Phi_+) \wedge \tilde f - \frac16 (\tilde f \lrcorner \Phi_+) \wedge
   \theta & =  0 \\
   - \lambda *(\Phi_{- 3} \wedge \theta)_{mnpq} + \frac12  *(\Phi_{- 3} \wedge
   d \Delta)_{mnpq} - \frac1{12} *(F \wedge V_{- 3})_{mnpq}   & \nonumber \\
   + \frac14 F_{[mn}{}^{rs} (\Phi_{- 3})_{pq]rs} + \frac16 (\theta
   \lrcorner \Phi_{- 3}) \wedge \tilde f - \frac16 (\tilde f \lrcorner \Phi_{- 3}) \wedge
   \theta & =  0 \\
   -2 \lambda (\Psi \wedge \theta)_{mnpq} + (\Psi \wedge d
   \Delta)_{mnpq} +\frac19 (* \Psi)_{[mnp}{}^{rst} F_{q]rst} & \nonumber \\
   + \frac23 K_{[m}{}^r F_{npq]r} + \frac13 *(\Psi \wedge \tilde f \wedge
   \theta)_{mnpq} + \frac13 (K \wedge \tilde f \wedge \theta)_{mnpq} & =  0
\end{align}
Notice that since we are using a set of generators $\{\gamma_A \xi_i\}$ instead of a basis, the equations above are not independent. In fact, as it is done explicitly for the special flux analysed in the paper, one only needs the constraints arising from contraction with $\xi_i$ and $\gamma_m \xi_i$.
%

\section{Killing properties of the Majorana-Weyl components}
\label{alflux}

Given that a background supports a covariantly constant Majorana
spinor $\xi$, the requirement for its  Majorana-Weyl
components to also solve the Killing spinor equation is
\begin{align}
[{\cal D}_m, \theta^r \gamma_r]\xi& =0 \;,\\
[Q, \theta^r \gamma_r]\xi & = 0 \;.
\end{align}
Here ${\cal D}_m$ and $Q$ are defined in (\ref{susyeq}), and we recall
that the eight-dimensional chirality matrix $\gamma_9$ was given in
terms of the 9-dimensional basis as $\gamma_9 = \theta^r \gamma_r$.
One can easily show that the commutators with $\gamma_9$ are expressed as
\begin{align}
  \left[A_m, \theta^r \gamma_r \right]& = 2 \lambda\left(\gamma_m -
    \theta_m \theta^n \gamma_n\right) + \frac{1}{12} F^{mnpq}\theta^r
  \gamma_{npqr} \;,\\
  \left[Q, \theta^r \gamma_r\right] & = \partial^n \Delta \theta^r
  \gamma_{nr} + \frac{1}{36}
  F^{mnpq}\theta_{[m}\gamma_{npq]}+\frac{2}{3} \tilde f ^m \theta^n
  \theta_{[m} \gamma_{n]} \;.
\end{align}
We now impose the orthogonality of the auxiliary direction $\theta$ on the fluxes and the fact that $\lambda=0$ for our specific choice of flux
\begin{align}
\left[A_m, \theta^r \gamma_r \right]& = \frac{1}{12} F^{mnpq}\theta^r \gamma_{npqr} \;,\\
\left[Q, \theta^r \gamma_r\right] & = \partial^n \Delta \theta^r \gamma_{nr} -\frac{1}{3} \tilde f ^m \gamma_m \;.
\end{align}
Let us examine further the first condition. For this we multiply it by
$\theta^s \gamma_s$ from the left and find after some simple gamma
matrix manipulations
\begin{align}
F^{mnpq} \gamma_{npq} \xi = 0\;,
\end{align}
where $\xi$ can be any (or both) of the Majorana spinors $\xi_{1,2}$,
which define our background. For definiteness we shall consider that
this equation is valid for $\xi = \xi_1$. Multiplying this from the
left with $\xi_2^T$ and using \eqref{bilinears} we obtain
\begin{equation}
  F^{mnpq} \Psi_{npq} = 0 \; .
\end{equation}
Using \eqref{Psipar}, the fact that for backgrounds which admit
varying parameter $\alpha$ the ninth direction $\theta$ is parallel to
$V_+$ and the independence of the flux on this ninth direction
($V_+$), we find
\begin{equation}
  \varphi \lrcorner F = 0 \; .
\end{equation}
This relation however can only be compatible with \eqref{dalpha} if
$F=0$ and therefore we conclude that for backgrounds in which the
parameter $\alpha$ varies, none of the Majorana--Weyl components can
satisfy by itself the Killing spinor equation.

\end{document}